\documentclass[11pt]{article}

\usepackage{tikz}
\usetikzlibrary{positioning,shapes,arrows.meta}

\usepackage{amssymb}  
\usepackage{textcomp} 
\usepackage{tcolorbox}

\usepackage{enumitem}
\usepackage{acl}
\usepackage{rotating}
\usepackage{multirow}
\usepackage{graphicx}
\usepackage{subcaption}
\usepackage{float}
\usepackage{booktabs} 
\usepackage{balance}  
\usepackage{amsmath}  

\usepackage{colortbl}
\usepackage{times}
\usepackage{latexsym}
\usepackage[english]{babel}

\usepackage[T1]{fontenc}
\usepackage{placeins}
\usepackage{stfloats}
\usepackage[utf8]{inputenc}

\usepackage{microtype}
\usepackage{cleveref}

\usepackage{listings}
\usepackage{tcolorbox}
\tcbuselibrary{breakable}

\newtcolorbox{promptbox}{
  colback=gray!8,
  colframe=gray!40,
  boxrule=0.5pt,
  arc=1mm,
  breakable,
  left=3pt,
  right=3pt,
  top=3pt,
  bottom=3pt,
  fontupper=\footnotesize
}

\usepackage{inconsolata}

\usepackage{caption}

\usepackage{graphicx}
\usepackage{mdframed} 

\title{RECOR: Reasoning-focused Multi-turn Conversational Retrieval Benchmark}

\author{
  Mohammed Ali\textsuperscript{1}\thanks{Corresponding author: \texttt{mohammed.ali@uibk.ac.at}} \quad
  Abdelrahman Abdallah\textsuperscript{1} \quad
  Amit Agarwal\textsuperscript{2} \quad
  Hitesh Laxmichand Patel\textsuperscript{2} \quad
  Adam Jatowt\textsuperscript{1} \\
  \\
  \textsuperscript{1}University of Innsbruck, Innsbruck, Austria \quad
  \textsuperscript{2}Oracle AI
}

\begin{document}
\maketitle


\begin{abstract}
Existing benchmarks treat multi-turn conversation and reasoning-intensive retrieval separately, yet real-world information seeking requires both. To bridge this gap, we present a benchmark for reasoning-based conversational information retrieval comprising 707 conversations (2,971 turns) across eleven domains. To ensure quality, our Decomposition-and-Verification framework transforms complex queries into fact-grounded multi-turn dialogues through multi-level validation, where atomic facts are verified against sources and explicit retrieval reasoning is generated for each turn. Comprehensive evaluation reveals that combining conversation history with reasoning doubles retrieval performance (Baseline .236 $\rightarrow$ History+Reasoning .479 nDCG@10), while reasoning-specialized models substantially outperform dense encoders. Despite these gains, further analysis highlights that implicit reasoning remains challenging, particularly when logical connections are not explicitly stated in the text. \footnote{\url{https://github.com/RECOR-Benchmark/RECOR}}
\end{abstract}

\section{Introduction}

Information seeking is inherently conversational. Exploring complex topics, users rarely employ single, comprehensive queries, instead engaging in multi-turn dialogue, progressively refining their understanding through iterative exchanges. This motivates Conversational Information Retrieval (CIR) systems that support ongoing interaction over isolated sessions.

Retrieval-Augmented Generation (RAG) has emerged as a critical approach for grounding LLM responses in external knowledge~\citep{lewis2020retrieval, chen2024benchmarking}. The primary focus of RAG benchmarks has been on single-turn interactions~\citep{friel2024ragbench, yang2024crag}. Multi-turn RAG, where each turn depends on preceding context, presents additional challenges that remain underexplored~\citep{katsis2025mtrag}.

A parallel challenge exists in retrieval itself. The BRIGHT benchmark~\citep{su2024bright} demonstrated that current retrieval systems struggle significantly with reasoning-intensive queries. These queries require multi-step inference, connecting disparate pieces of information, and drawing non-obvious conclusions. Standard lexical and semantic matching fails when relevance depends on logical reasoning rather than surface similarity.

These two challenges, multi-turn conversation and reasoning-intensive retrieval, have been studied separately. However, no existing benchmark combines them. Yet real information needs often require both. A researcher investigating climate policy might ask why carbon prices worked in some countries but not others, then whether the successful cases shared common features, then how those lessons apply to developing nations, etc. Each turn requires reasoning to identify relevant documents, and each answer builds on previous understanding.

Bridging this gap requires high-quality multi-turn conversations, yet these are difficult to obtain. Human annotation is expensive and does not scale. Synthetic generation is fast but often produces shallow conversations with repetitive content and limited document diversity. Without principled methods that incorporate explicit reasoning, one cannot create the benchmarks needed to advance conversational search systems.

We address this challenge through a Decomposition-and-Verification framework that transforms complex single-turn queries into grounded multi-turn dialogues. Given a query with its gold answer and supporting documents, the system validates document-answer alignment, then decomposes the answer into granular aspects representing distinct information facets. From each aspect, atomic facts are extracted and verified against sources. For each verified aspect, the system generates a focused sub-question, scores relevant documents, and produces a grounded conversational turn.

Unlike approaches relying solely on semantic matching, our framework generates explicit retrieval reasoning for each turn, specifying what information documents should contain and what signals indicate relevance. This reasoning-guided selection produces natural document diversity, with different turns drawing upon different passages.

Our contributions: (1) A Decomposition-and-Verification framework that transforms complex queries into fact-grounded multi-turn dialogues, reusable for conversational data synthesis; (2) Using this framework, a benchmark for reasoning-intensive conversational retrieval across eleven domains (707 conversations, 2,971 turns); (3) Extensive experiments on retrieval and generation showing that history + reasoning doubles retrieval (.236 $\rightarrow$ .479 nDCG@10), reasoning-specialized models substantially outperform dense encoders, and implicit reasoning remains challenging.


\section{Related Work}

We summarize key related work below while Appendix~\ref{sec:extended_related_work} provides a comprehensive review.

\paragraph{Conversational Search.} Researchers have developed several benchmarks for multi-turn information seeking. QReCC \cite{anantha2021open} contains 14K open-domain conversations where users ask follow-up questions that require resolving references to previous turns. TREC iKAT \cite{aliannejadi2024trec} extends this setting by incorporating user personas across 20 topics, requiring systems to personalize responses based on user background. MTRAG \cite{katsis2025mtrag} evaluates multi-turn RAG systems on 110 conversations with challenges like unanswerable questions across four domains. While capturing conversational complexity, these benchmarks assume document relevance relies primarily on semantic similarity.

\paragraph{Reasoning for Retrieval.} BRIGHT \cite{su2024bright} showed that current retrievers fail on queries requiring reasoning. Systems achieving 59.0 nDCG@10 on standard benchmarks score only 18.3 on BRIGHT. The benchmark also showed that providing chain-of-thought reasoning improves retrieval by up to 12.2 points. However, BRIGHT only covers single turn queries (Table~\ref{tab:benchmark_comparison}).

\begin{table}[t]
\centering

\small
\setlength{\tabcolsep}{4pt}
\begin{tabular}{lccccc}
\toprule
\textbf{Benchmark} & \textbf{MT} & \textbf{RI} & \textbf{FG} & \textbf{RR} & \textbf{Dom.} \\
\midrule
QReCC & \checkmark & \texttimes & \texttimes & \texttimes & — \\
TREC iKAT & \checkmark & \texttimes & \texttimes & \texttimes & 20$^*$ \\
MTRAG & \checkmark & \texttimes & \texttimes & \texttimes & 4 \\
BRIGHT & \texttimes & \checkmark & \texttimes & \checkmark & 12 \\
\midrule
\textbf{Ours} & \checkmark & \checkmark & \checkmark & \checkmark & \textbf{11} \\
\bottomrule
\end{tabular}
\caption{Benchmark comparison. MT: Multi-turn. RI: Reasoning-intensive. FG: Fact-grounded. RR: Retrieval reasoning. Dom.: Number of domains. QReCC is open-domain. $^*$TREC iKAT has 20 topics, not domains.}
\label{tab:benchmark_comparison}
\end{table}

\section{Dataset Construction and Analysis}
\label{sec:dataset}

Figure~\ref{fig:pipeline} illustrates our complete pipeline, which we detail in the following subsections. See Appendix~\ref{appendix:pipeline} for a detailed example. (Pipeline prompts: see Appendix~\ref{appendix:prompts})

\begin{figure*}[t]
\centering
\includegraphics[width=.8\textwidth]{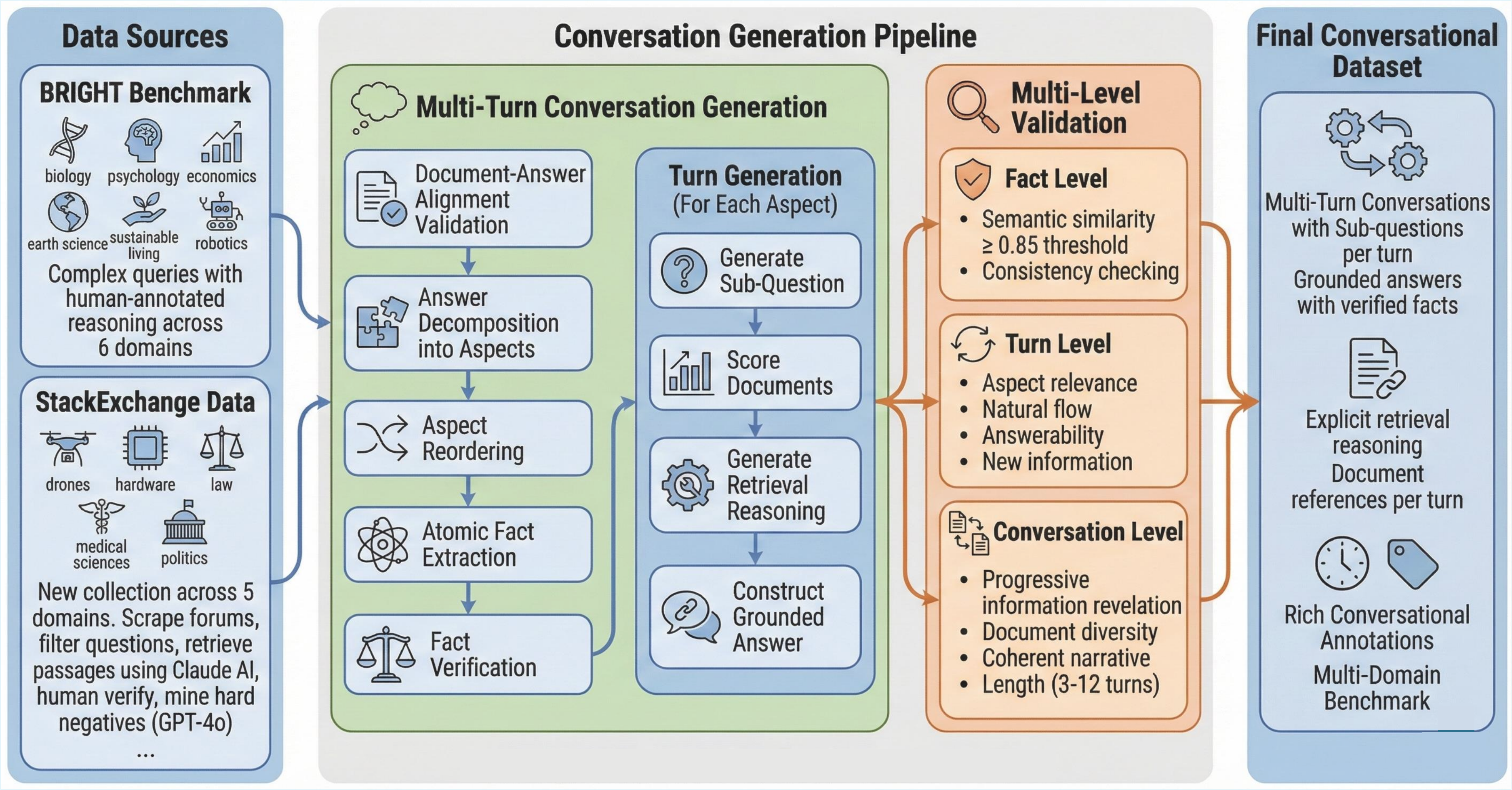}
\caption{Overview of Decomposition-and-Verification framework for generating grounded multi-turn conversations.}
\label{fig:pipeline}
\end{figure*}

\subsection{Data Sources and Collection}

Our benchmark draws from two complementary sources: the existing BRIGHT benchmark and the newly collected StackExchange data. This combination provides both established reasoning-intensive queries and domain-specific technical questions, enabling comprehensive evaluation across eleven domains.



\subsubsection{BRIGHT Benchmark}
We leverage the BRIGHT dataset~\citep{su2024bright} as a foundational component, utilizing its complex queries and human-annotated reasoning across six domains: biology, psychology, economics, earth science, sustainable living, and robotics. These queries require logical inference rather than surface similarity, and the original positive and negative passages, curated for multi-step reasoning, are preserved.

\subsubsection{StackExchange Collection}

To extend domain coverage, we collected additional complex queries from StackExchange across five complementary domains: drones, hardware, law, medical sciences, and politics. Our collection methodology ensures data quality comparable to BRIGHT.

\paragraph{Question-Answer Extraction.} We scraped StackExchange domain-specific forums, applying strict selection criteria. Questions must be sufficiently complex and detailed, filtering out simple requests or yes/no questions. Only questions with verified (accepted) answers were retained, and answer completeness was verified to ensure meaningful responses requiring reasoning.

\paragraph{Positive Passage Retrieval.} For each question-answer pair, we employed a two-stage process. First, Claude AI Assistant retrieved candidate passages from diverse sources (Wikipedia, technical blogs, research articles). The assistant was prompted to find passages providing information necessary to answer the question, even if not directly stating the answer. Second, human annotators verified each retrieved passage for relevance and quality consistency across domains.

\paragraph{Hard Negative Mining.} Standard negative sampling (e.g., BM25 top-k) produces passages easily distinguishable from positives. We developed a query-based approach where GPT-4o \citep{openai2024gpt4o} generates topically related queries focused on unhelpful aspects. We retrieved web pages via Google Custom Search API, excluding original positive sources, and segmented them into comparable-length passages. Automated verification removed duplicates and passages with high lexical overlap. This method produces hard negatives that share domain vocabulary but lack the specific information needed for answering, creating a genuine challenge for reasoning-based retrieval.



\subsection{Multi-Turn Conversation Generation}

Transforming single-turn QA pairs into multi-turn conversations requires balancing document grounding with natural dialogue flow. Our Decomposition-and-Verification framework achieves this through systematic answer decomposition, fact-level verification, and explicit retrieval reasoning.

\subsubsection{Answer Decomposition and Fact Extraction}

Given a question $Q$, answer $A$, and supporting documents $D = \{d_1, d_2, \ldots, d_n\}$, our pipeline first validates document-answer alignment, filtering irrelevant or weakly related passages.

The validated answer is decomposed into granular aspects: distinct components representing coherent subtopics or reasoning steps. Aspect extraction is iterative, checking each candidate for meaningful content, non-overlapping, and sufficient document coverage. For example, a climate policy answer might yield aspects covering carbon pricing, implementation challenges, and economic trade-offs.

Aspects are reordered to ensure natural conversational flow (e.g., progressing from concepts to implications; see Appendix~\ref{appendix:prompts} for prompts). From each aspect, we extract \textbf{atomic facts}: minimal statements that cannot be divided further. Each fact is verified against source documents via semantic similarity and logical consistency; facts without clear support are discarded.





\subsubsection{Turn Generation with Retrieval Reasoning}

For each verified aspect, the system generates a focused conversational turn. A natural follow-up question targets the specific aspect, building upon preceding context while introducing new information needs. Relevant documents are scored on multiple criteria: coverage of atomic facts, answer completeness, explanation clarity, and penalties for misleading content. Unlike semantic similarity alone, this scoring considers whether documents contain the specific information needed to address the sub-question.

For each selected document, we generate explicit reasoning explaining its relevance: what information it contains and what signals indicate its usefulness. The turn's answer is constructed from verified facts with direct connections to source passages. This reasoning-guided approach produces natural document diversity across turns, as different aspects draw upon different information subsets.




\subsubsection{Multi-Level Validation}

Quality assurance operates at three levels. \textbf{At the fact level}, each atomic fact is verified against source documents, and facts are accepted if they are explicitly stated, clearly implied, or are reasonable paraphrases. \textbf{At the turn level}, generated sub-questions are validated for aspect relevance, natural flow, answerability from available documents, and distinctiveness from previous turns. \textbf{At the conversation level}, complete dialogues are evaluated for logical progression, document diversity across turns, coherent narrative structure, and appropriate length (3--12 turns). (See Appendix~\ref{appendix:prompts} for prompts).

\subsection{Dataset Statistics and Characteristics}

Our benchmark comprises 707 conversations with 2,971 turns across eleven domains (Table~\ref{tab:domain_stats}). The StackExchange subset contributes 220 conversations (956 turns) while BRIGHT provides 487 conversations (2,015 turns). Conversations average 4.20 turns (range 3--12) with 5.87 semantic aspects and 2.01 supporting documents per turn. The benchmark contains 507,141 total documents (2,900 positive, 504,241 hard negatives), creating a challenging 174:1 negative-to-positive ratio. Queries average 18 words, answers average 70 words, and conversation history accumulates to approximately 155 words. The eleven domains cover diverse reasoning patterns: evidence-based inference (scientific), procedural reasoning (technical), and causal analysis (social). Detailed statistics are provided in Appendix~\ref{app:detailed_stats}.

\begin{table}[t]
\centering
\scriptsize
\begin{tabular}{lrrrrr}
\toprule
\textbf{Domain} & \textbf{\#Conv.} & \textbf{\#Turns} & \textbf{Avg \#T} & \textbf{Avg \#A} & \textbf{Avg \#D} \\
\midrule
Drones & 37 & 142 & 3.84 & 4.76 & 2.36 \\
Hardware & 46 & 188 & 4.09 & 5.20 & 2.10 \\
Law & 50 & 230 & 4.60 & 7.00 & 2.55 \\
Medical Sci. & 44 & 183 & 4.16 & 6.18 & 2.23 \\
Politics & 43 & 213 & 4.95 & 7.42 & 2.49 \\
Biology & 85 & 362 & 4.26 & 5.89 & 1.56 \\
Earth Sci. & 98 & 454 & 4.63 & 6.23 & 1.58 \\
Economics & 74 & 288 & 3.89 & 5.62 & 2.28 \\
Psychology & 84 & 333 & 3.96 & 5.65 & 2.16 \\
Robotics & 68 & 259 & 3.81 & 4.94 & 1.76 \\
Sust. Living & 78 & 319 & 4.09 & 5.79 & 1.88 \\
\midrule
\textbf{Total} & \textbf{707} & \textbf{2,971} & \textbf{4.20} & \textbf{5.87} & \textbf{2.01} \\
\bottomrule
\end{tabular}
\caption{Per-domain statistics. Avg \#T: average turns per conversation; Avg \#A: average aspects per conversation; Avg \#D: average documents per turn.}
\label{tab:domain_stats}
\end{table}

\subsection{Human Evaluation}
\label{sec:human_evaluation}

To validate conversation quality, three PhD students evaluated 200 randomly sampled conversations (balanced across domains) on four criteria using a 1--5 Likert scale: naturalness (dialogue flow), turn coherence (logical progression), question quality (meaningful sub-questions), and groundedness (answer-document alignment). Additionally, annotators manually revised the automatically selected documents and generated retrieval reasoning for each turn, correcting any misalignments or inaccuracies. Results show strong performance: naturalness 4.2 ($\kappa$=0.79), turn coherence 4.6 ($\kappa$=0.77), question quality 4.2 ($\kappa$=0.75), and groundedness 4.4 ($\kappa$=0.73), validating our Decomposition-and-Verification framework.

To extend this validation to the full dataset, we used GPT-4o to automatically assess all conversations on the same four criteria. LLM scores align closely with human judgments, with differences ranging from +0.14 to +0.35, all within acceptable thresholds. We also analyzed turn-level dependency types and question patterns. Full results and evaluation prompts are provided in Appendix~\ref{appendix:automatic_validation}.

\section{Experiments}
\label{sec:experiments}

We evaluate our benchmark through retrieval experiments that assess passage retrieval quality (\S\ref{sec:retrieval_eval}) and generation experiments that judge answer quality from retrieved passages (\S\ref{sec:generation_eval}).

\subsection{Retrieval Evaluation}
\label{sec:retrieval_eval}

\subsubsection{Setup}
\label{sec:retrieval_setup}

\paragraph{Query Processing Strategies.}

Multi-turn queries challenge retrieval systems through coreferences and context dependencies. We evaluate five strategies: \textbf{Baseline} uses the raw current query. \textbf{Query Rewrite} employs GPT-4o to generate self-contained queries. \textbf{Reasoning} decomposes information needs via explicit search rationale. \textbf{History} prepends full conversation history. \textbf{History+Reasoning} combines context with reasoning.



\paragraph{Retrieval Models.}
We evaluate eight retrievers spanning three categories. Reasoning-specialized models include DIVER \citep{wu2024diver} and ReasonIR \citep{shao2024reasonir}, trained on multi-hop queries requiring inference beyond surface matching. General-purpose dense encoders include BGE \citep{xiao2024bge}, E5 \citep{wang2024e5}, Contriever \citep{izacard2022contriever}, SFR \citep{meng2024sfr}, and Qwen \citep{bai2023qwen}. For lexical retrieval, we include BM25 \citep{robertson2009bm25} as a sparse baseline.


\paragraph{Metrics.}
We report nDCG@10 \citep{jarvelin2002ndcg} as our primary metric. For each retriever and strategy combination, we compute nDCG@10 per turn, an average within each domain, then macro-average across all 11 domains to ensure equal weight regardless of domain size. Statistical tests use paired comparisons across 23,768 turn-level observations. Full metrics appear in Appendix~\ref{sec:app_metrics}.

\subsubsection{Results}
\label{sec:retrieval_results}


Table~\ref{tab:main_results} presents retrieval performance across all configurations. The average Baseline score (.236) confirms that retrieving relevant passages is difficult without context, as later-turn queries contain implicit references requiring prior conversational knowledge. History+Reasoning achieves .479, doubling Baseline performance (paired $t=119.67$, $p<0.001$).

Reasoning-specialized retrievers consistently outperform other architectures. DIVER achieves the highest score of .584 with History+Reasoning (95\% CI: .520 to .649). ReasonIR reaches .552 (95\% CI: .501 to .603). Both substantially exceed the best dense encoder, SFR at .464, confirming that models trained for multi-hop reasoning transfer effectively to conversational retrieval. Complete statistical details are provided in Appendix~\ref{sec:app_stats}.

\begin{table}[t]
\centering

\small
\resizebox{\columnwidth}{!}{%
\begin{tabular}{@{}l ccccc@{}}
\toprule
\textbf{Retriever} & \textbf{Base} & \textbf{QR} & \textbf{Reas} & \textbf{Hist} & \textbf{H+R} \\
\midrule
\rowcolor{gray!15} \multicolumn{6}{c}{\textbf{\textit{Reasoning-specialized}}} \\
\midrule
\textbf{DIVER} & .347 & .430 & .496 & .545 & \textbf{.584} \\
\textbf{ReasonIR} & .266 & .357 & .494 & .496 & .552 \\
\midrule
\rowcolor{gray!15} \multicolumn{6}{c}{\textbf{\textit{Dense encoders}}} \\
\midrule
\textbf{Qwen} & .269 & .345 & .399 & .425 & .461 \\
\textbf{SFR} & .240 & .324 & .396 & .429 & .464 \\
\textbf{BGE} & .230 & .328 & .347 & .411 & .445 \\
\textbf{E5} & .183 & .272 & .352 & .404 & .429 \\
\textbf{Contriever} & .168 & .232 & .303 & .366 & .409 \\
\midrule
\rowcolor{gray!15} \multicolumn{6}{c}{\textbf{\textit{Lexical}}} \\
\midrule
\textbf{BM25} & .185 & .288 & .360 & .446 & .489 \\
\midrule
\textbf{Average} & .236 & .322 & .393 & .440 & \textbf{.479} \\
\textbf{95\% CI} & {\scriptsize$\pm$.004} & {\scriptsize$\pm$.004} & {\scriptsize$\pm$.004} & {\scriptsize$\pm$.004} & {\scriptsize$\pm$.004} \\
\bottomrule
\end{tabular}%
}
\caption{Retrieval performance (nDCG@10, macro-averaged across 11 domains). Base=Baseline, QR=Query Rewrite, Reas=Reasoning, Hist=History, H+R=History+Reasoning. All strategy differences are significant ($p<0.001$, paired $t$-test, $n$=23,768).}
\label{tab:main_results}
\end{table}

\subsubsection{Ablation Study}
\label{sec:component}


A central question is how much each component contributes to retrieval improvement. Table~\ref{tab:ablation} presents an ablation isolating the individual and combined effects of History and Reasoning.

Reasoning alone improves retrieval by 67\% over Baseline, helping retrievers identify passages requiring inference rather than simple keyword matching. History provides larger gains (+86\%) by directly resolving coreferences and ellipsis that make later-turn queries incomplete. Combined, History+Reasoning yields +103\% improvement, larger than either component alone but smaller than their sum (153\%), confirming that both address some overlapping challenges while each contributes unique value.

The pattern varies across architectures. ReasonIR benefits equally from both components (+86\% each), consistent with its reasoning-intensive training. BM25 benefits most from History (+141\%) because lexical matching needs additional query terms. Dense encoders show balanced gains, with SFR improving +65\% from Reasoning and +79\% from History. Notably, all retrievers benefit when both components are combined.

\begin{table}[t]
\centering

\small
\setlength{\tabcolsep}{2.5pt}
\begin{tabular}{l|cccc}
\toprule
\textbf{Retriever} & \textbf{Base} & \textbf{+Reas} & \textbf{+Hist} & \textbf{H+R} \\
\midrule
DIVER & .347 & .496 (+43\%) & .545 (+57\%) & .584 (+68\%) \\
ReasonIR & .266 & .494 (+86\%) & .496 (+86\%) & .552 (+108\%) \\
BM25 & .185 & .360 (+95\%) & .446 (+141\%) & .489 (+164\%) \\
Qwen & .269 & .399 (+48\%) & .425 (+58\%) & .461 (+71\%) \\
SFR & .240 & .396 (+65\%) & .429 (+79\%) & .464 (+93\%) \\
\midrule
\textbf{Average} & .236 & .393 (+67\%) & .440 (+86\%) & .479 (+103\%) \\
\bottomrule
\end{tabular}
\caption{Ablation of Reasoning and History components (nDCG@10, 11 domains). Parentheses: \% improvement over Baseline.}
\label{tab:ablation}
\end{table}



\subsubsection{Turn Position Effects}
\label{sec:turn_position}

\begin{figure}[t]
    \centering
    
    \includegraphics[width=.7\columnwidth]{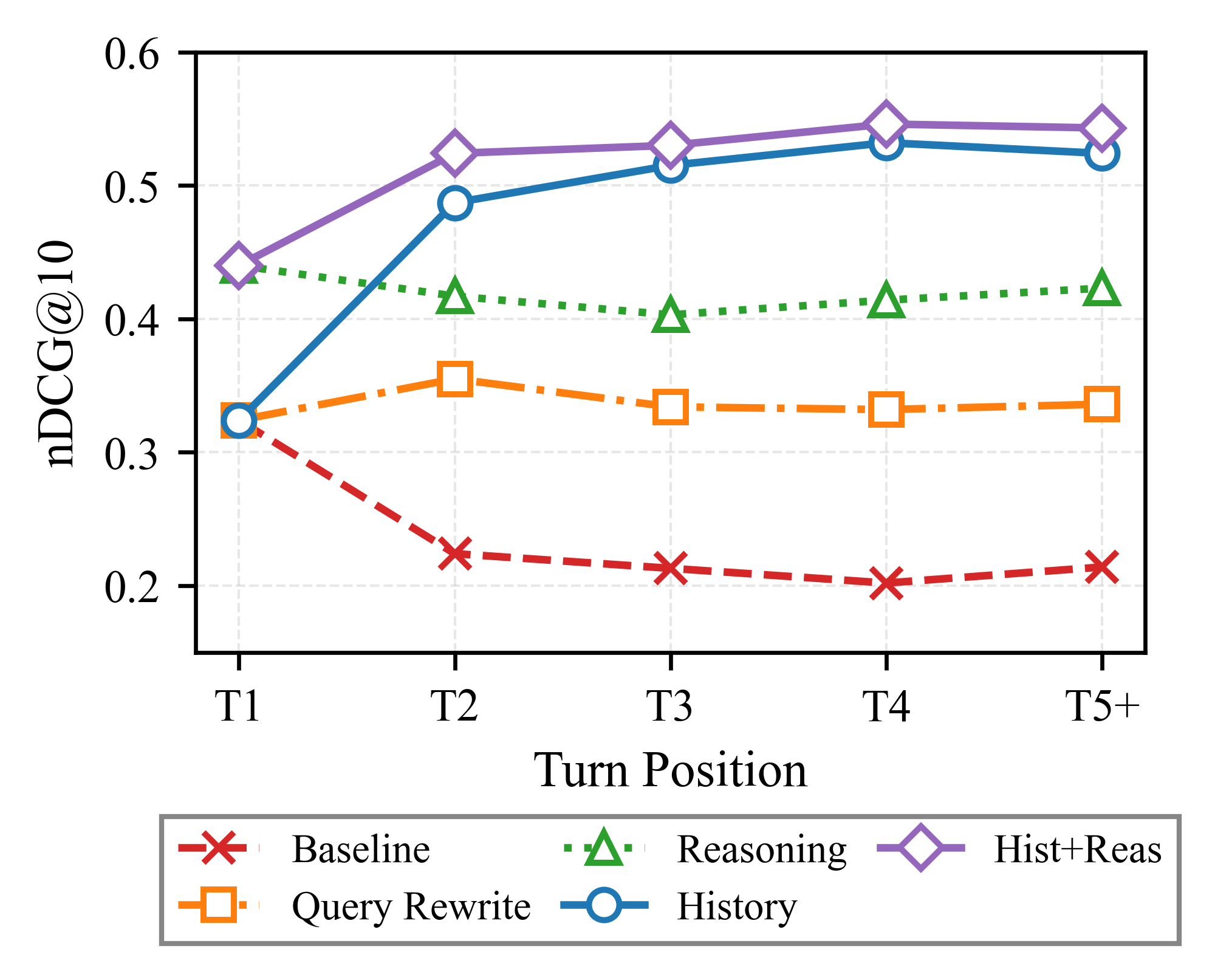}
    \caption{Retrieval performance (nDCG@10) by turn position (avg. 8 retrievers, 11 domains). Non-reasoning strategies share equal T1 performance; divergence emerges from Turn 2 as context dependence grows.}
    \label{fig:turn_position}
\end{figure}

Figure~\ref{fig:turn_position} illustrates retrieval performance evolution across conversation turns, aggregated across all retrievers and domains. At Turn 1, queries are self-contained. Baseline, Query Rewrite, and History produce identical scores (.324) since no history exists. However, Reasoning-based strategies show higher T1 performance (.440) because reasoning is appended even at Turn 1, providing additional retrieval signals. From Turn 2 onward, strategy differences emerge as queries become context-dependent. Baseline performance drops sharply to .224 at Turn 2 (31\% decline) and stabilizes around .20, confirming that later-turn queries cannot be interpreted in isolation.

In contrast, History-based strategies improve as conversations progress, rising from .487 at Turn 2 to .532 at Turn 4. The performance gap over Baseline widens from 117\% to 163\%, confirming that the benchmark captures genuine conversational complexity where accumulated context is increasingly critical. Per-retriever analysis is provided in Appendix~\ref{sec:app_turn}.

\subsubsection{Domain Variation}
\label{sec:domain}

Retrieval difficulty varies across subject areas (Table~\ref{tab:domain}). Life science domains (Psychology, Biology) achieve the highest scores (.733, .732) due to standardized terminology, while technical domains (Robotics, Hardware) show lower absolute performance but larger relative improvements (+201\%, +80\%), suggesting specialized vocabulary creates baseline difficulty that History+Reasoning addresses.

DIVER performs best in 7 of 11 domains, while ReasonIR excels in technical and legal domains (Law, Hardware, Drones) requiring procedural reasoning. BM25 achieves the highest Robotics score (.587) as specific technical terms favor exact matching. See Appendix~\ref{sec:app_domain} for the complete retriever-by-domain matrix.


\begin{table}[t]
\centering

\small
\setlength{\tabcolsep}{3pt}
\begin{tabular}{l|cccl}
\toprule
\textbf{Domain} & \textbf{Base} & \textbf{H+R} & \textbf{$\Delta$} & \textbf{Best} \\
\midrule
Psychology & .391 & .733 & +87\% & DIVER \\
Biology & .450 & .732 & +63\% & DIVER \\
Earth Science & .428 & .694 & +62\% & DIVER \\
Sust.\ Living & .405 & .682 & +68\% & DIVER \\
Robotics & .195 & .587 & +201\% & BM25 \\
Politics & .362 & .582 & +61\% & DIVER \\
Economics & .283 & .572 & +102\% & DIVER \\
Law & .331 & .540 & +63\% & ReasonIR \\
Drones & .325 & .499 & +53\% & ReasonIR \\
Medical Sci. & .395 & .488 & +24\% & DIVER \\
Hardware & .253 & .455 & +80\% & ReasonIR \\
\bottomrule
\end{tabular}
\caption{Domain results (nDCG@10). Base: DIVER Baseline (strongest retriever). H+R: best retriever with History+Reasoning. $\Delta$: relative improvement. Best: top retriever per domain.}
\label{tab:domain}
\end{table}



\subsubsection{Conversation Complexity}
\label{sec:complexity}

Each conversation in our benchmark covers multiple semantic aspects derived from decomposing the source answer into distinct subtopics. For example, a sustainability conversation with five aspects might discuss carbon footprint calculation, renewable energy alternatives, policy incentives, implementation barriers, and long-term environmental impact as separate threads. We examined whether complex conversations with more aspects would benefit most from History+Reasoning (Table~\ref{tab:complexity}).

\begin{table}[t]
\centering

\small
\setlength{\tabcolsep}{4pt}
\begin{tabular}{l|cccc}
\toprule
\textbf{Complexity} & \textbf{Turns} & \textbf{Base} & \textbf{H+R} & \textbf{$\Delta$} \\
\midrule
Low (3--4 aspects) & 463 & .428 & .813 & +90\% \\
Medium (5--6 aspects) & 734 & .454 & .810 & +78\% \\
High (7+ aspects) & 1,067 & .509 & .810 & +59\% \\
\bottomrule
\end{tabular}
\caption{Retrieval performance by conversation complexity (queries from Turn 2 onward, maximum nDCG@10 across all retrievers). Low-complexity conversations benefit more from History+Reasoning than high-complexity ones.}
\label{tab:complexity}
\end{table}

The results reveal a counterintuitive pattern: low-complexity conversations benefit most from History+Reasoning (+90\%) while high-complexity ones benefit least (+59\%). Complex conversations contain richer content across turns (more entities, terminology, and concepts), giving Baseline more signals to work with, even without history. This explains why Baseline performs better on complex conversations (.509 vs.\ .428 for simple ones), while History+Reasoning reaches a similar ceiling ($\sim$.81) regardless of complexity. It also aligns with prior finding that context augmentation helps less when queries already contain rich information~\citep{krasakis2024contextualizing}.

\subsubsection{Failure Analysis}
\label{sec:failure}

We categorize all 2,971 queries by the maximum nDCG@10 achieved by any retriever with History+Reasoning. Results show 63.2\% of queries achieve nDCG@10 $\geq$ 0.8, demonstrating that combining history with reasoning addresses most conversational retrieval challenges. Another 22.4\% score between 0.5 and 0.8, while 8.1\% score between 0.3 and 0.5. The remaining 6.3\% (187 queries) are persistent failures where no configuration reaches 0.3.


Among these failures, Turn 1 queries appear more often than expected (37\% of failures vs.\ 24\% of all queries). This reveals two failure types: Turn 1 failures are inherently difficult questions where context cannot help, while later-turn failures (63\%) occur when history contains relevant information but retrieval still fails.

Manual inspection reveals two primary failure modes. First, informal phrasing creates vocabulary mismatch. For example, ``Is extra cooling really necessary for the 10700K?'' uses colloquial terms while documents discuss ``Intel Core i7-10700K thermal design power specifications.'' Second, implicit domain knowledge is required. For example, ``Does that shift have something to do with how the red and green cones balance each other?'' requires understanding that ``that shift'' refers to the Purkinje effect from earlier turns. Even with complete history, retrievers fail because the reasoning chain connecting these concepts is never explicitly stated. These failures suggest the need for retrieval systems capable of implicit reasoning.

\subsection{Generation Evaluation}
\label{sec:generation_eval}

\subsubsection{Setup}

\paragraph{Generators.} We evaluate seven instruction-tuned language models spanning three size categories: large (Llama-3.3-70B, Qwen2.5-72B), medium (Gemma-3-12B, Qwen2.5-14B), and small (Llama-3.1-8B, Qwen2.5-7B, Mistral-7B).

\paragraph{Retrieval Modes.} We evaluate three conditions. \textbf{Oracle} provides gold-standard passages containing the correct answer, establishing an upper bound. \textbf{Retrieved} uses passages from our best retrieval configuration (DIVER with History+Reasoning, $k$=5). \textbf{NoRetrieval} generates answers without retrieved passages, relying solely on the model's parametric knowledge.

\paragraph{Evaluation.} We use GPT-4o as a judge to score five dimensions on a 1--5 scale: Correctness, Completeness, Relevance, Coherence, and Faithfulness. Faithfulness evaluates whether answers are grounded in retrieved passages and applies only to Oracle and Retrieved modes. Following \S\ref{sec:retrieval_setup}, scores are normalized to 0--1 and macro-averaged across domains. Lexical metrics in Appendix~\ref{sec:app_lexical}.

\subsubsection{Results}

Table~\ref{tab:gen_domain} presents generation quality across domains and models using Retrieved mode. Complete per-domain results for all modes appear in Appendix~\ref{sec:app_gen_domain}.

\begin{table}[t]
\centering

\small
\resizebox{\columnwidth}{!}{%
\begin{tabular}{@{}l ccccccc@{}}
\toprule
\textbf{Domain} & \rotatebox{90}{\textbf{Llama-70B}} & \rotatebox{90}{\textbf{Qwen-72B}} & \rotatebox{90}{\textbf{Gemma-12B}} & \rotatebox{90}{\textbf{Qwen-14B}} & \rotatebox{90}{\textbf{Llama-8B}} & \rotatebox{90}{\textbf{Qwen-7B}} & \rotatebox{90}{\textbf{Mistral-7B}} \\
\midrule
\rowcolor{gray!15} \multicolumn{8}{c}{\textbf{\textit{Life Sciences}}} \\
\midrule
\textbf{Psychology} & \textbf{.900} & .839 & .852 & .862 & .843 & .809 & .841 \\
\textbf{Biology} & \textbf{.897} & .858 & .840 & .793 & .839 & .847 & .820 \\
\textbf{Medical Sci.} & \textbf{.887} & .843 & .844 & .851 & .840 & .824 & .811 \\
\midrule
\rowcolor{gray!15} \multicolumn{8}{c}{\textbf{\textit{Physical Sciences}}} \\
\midrule
\textbf{Earth Sci.} & .767 & \textbf{.785} & .726 & .678 & .756 & .754 & .713 \\
\midrule
\rowcolor{gray!15} \multicolumn{8}{c}{\textbf{\textit{Technical}}} \\
\midrule
\textbf{Hardware} & \textbf{.890} & .843 & .842 & .833 & .819 & .823 & .808 \\
\textbf{Robotics} & \textbf{.868} & .847 & .822 & .842 & .809 & .827 & .785 \\
\textbf{Drones} & .873 & .836 & \textbf{.864} & .835 & .842 & .818 & .806 \\
\midrule
\rowcolor{gray!15} \multicolumn{8}{c}{\textbf{\textit{Social Sciences}}} \\
\midrule
\textbf{Economics} & \textbf{.895} & .871 & .861 & .852 & .845 & .849 & .839 \\
\textbf{Law} & \textbf{.879} & .845 & .842 & .844 & .825 & .812 & .811 \\
\textbf{Politics} & \textbf{.858} & .841 & .845 & .834 & .813 & .805 & .804 \\
\midrule
\rowcolor{gray!15} \multicolumn{8}{c}{\textbf{\textit{Applied}}} \\
\midrule
\textbf{Sust.\ Living} & \textbf{.888} & .846 & .849 & .845 & .838 & .823 & .817 \\
\midrule
\textbf{Average} & \textbf{.880} & .850 & .835 & .824 & .833 & .817 & .805 \\
\bottomrule
\end{tabular}%
}
\caption{Generation quality by domain (Retrieved mode). LLM-Judge average score. Bold indicates best model per domain.}
\label{tab:gen_domain}
\end{table}

Llama-3.3-70B achieves the highest average score (.880) and leads in 9 of 11 domains. This dominance reflects its larger training corpus and stronger instruction-following capabilities for synthesizing retrieved information.

Life Sciences domains achieve the highest scores (.887--.900), as these domains feature well-structured factual content with clear answer patterns. In contrast, Earth Science proves most challenging (.678--.785), where complex spatiotemporal reasoning about geological processes remains difficult for current LLMs.

Small models (7--8B) score only 5--7 points below 70B+ models on average. This gap is notably smaller than in retrieval tasks (\S\ref{sec:retrieval_results}), suggesting that retrieved passages reduce the dependency on model scale by providing external knowledge that compensates for smaller models' limited parametric capacity.

\subsubsection{Retrieval Mode Analysis}

Table~\ref{tab:gen_mode} compares generation quality across retrieval conditions.

\begin{table}[t]
\centering

\small
\setlength{\tabcolsep}{4pt}
\begin{tabular}{@{}l ccc@{}}
\toprule
\textbf{Model} & \textbf{Oracle} & \textbf{Retrieved} & \textbf{No.Retr.} \\
\midrule
\rowcolor{gray!15} \multicolumn{4}{c}{\textbf{\textit{Large}}} \\
\midrule
\textbf{Llama-3.3-70B} & \textbf{.920} & \textbf{.888} & \textbf{.859} \\
\textbf{Qwen2.5-72B} & .876 & .860 & .837 \\
\midrule
\rowcolor{gray!15} \multicolumn{4}{c}{\textbf{\textit{Medium}}} \\
\midrule
\textbf{Gemma-3-12B} & .890 & .854 & .824 \\
\textbf{Qwen2.5-14B} & .874 & .852 & .824 \\
\midrule
\rowcolor{gray!15} \multicolumn{4}{c}{\textbf{\textit{Small}}} \\
\midrule
\textbf{Llama-3.1-8B} & .862 & .840 & .789 \\
\textbf{Qwen2.5-7B} & .849 & .835 & .803 \\
\textbf{Mistral-7B} & .849 & .819 & .779 \\
\midrule
\textbf{Average} & \textbf{.874} & \textbf{.850} & \textbf{.816} \\
\bottomrule
\end{tabular}
\caption{Generation quality by retrieval mode. LLM-Judge scores macro-averaged across 11 domains. NoRetrieval excludes Faithfulness.}
\label{tab:gen_mode}
\end{table}


Oracle achieves .874 average, confirming that LLMs generate high-quality answers when given relevant context. The gap between Oracle and Retrieved (2.4 points) represents the cost of imperfect retrieval, while the gap between Retrieved and NoRetrieval (3.4 points) confirms that RAG improves answer quality. The RAG benefit varies by model size: smaller models gain more (Llama-3.1-8B: +5.1 points) than larger models (Llama-3.3-70B: +2.9 points), because smaller models have weaker parametric knowledge and thus rely more on external retrieval.

\subsubsection{Fine-grained Analysis}

\paragraph{Retrieval-generation correlation.} 

\begin{table}[t]
\centering

\small
\setlength{\tabcolsep}{4pt}
\begin{tabular}{l|cc}
\toprule
\textbf{Domain} & \textbf{nDCG@10} & \textbf{LLM-Judge} \\
\midrule
Psychology & .733 & .856 \\
Biology & .732 & .856 \\
Economics & .572 & .867 \\
Sust.\ Living & .682 & .858 \\
Medical Sci. & .488 & .857 \\
Hardware & .455 & .851 \\
Law & .540 & .851 \\
Drones & .499 & .848 \\
Robotics & .587 & .843 \\
Politics & .582 & .843 \\
Earth Science & .694 & .740 \\
\bottomrule
\end{tabular}
\caption{Retrieval-generation correlation by domain. nDCG@10 from DIVER H+R. LLM-Judge averaged across all 7 generators. Pearson $r$ = .42.}
\label{tab:ret_gen_corr}
\end{table}

Table~\ref{tab:ret_gen_corr} examines whether better retrieval leads to better generation. Results show moderate correlation ($r$ = .42): high-retrieval domains (Psychology, Biology) achieve high generation scores, but the relationship is not deterministic. Hardware achieves .851 generation despite low retrieval (.455), while Earth Science scores only .740 despite high retrieval (.694). This aligns with findings from BRIGHT \cite{su2024bright}, where generation quality does not always reflect retrieval performance due to the generator's varying ability to integrate retrieved content across domains.


\paragraph{Turn position effects.} 




\begin{figure}[t]
    \centering
    \includegraphics[width=.7\columnwidth]{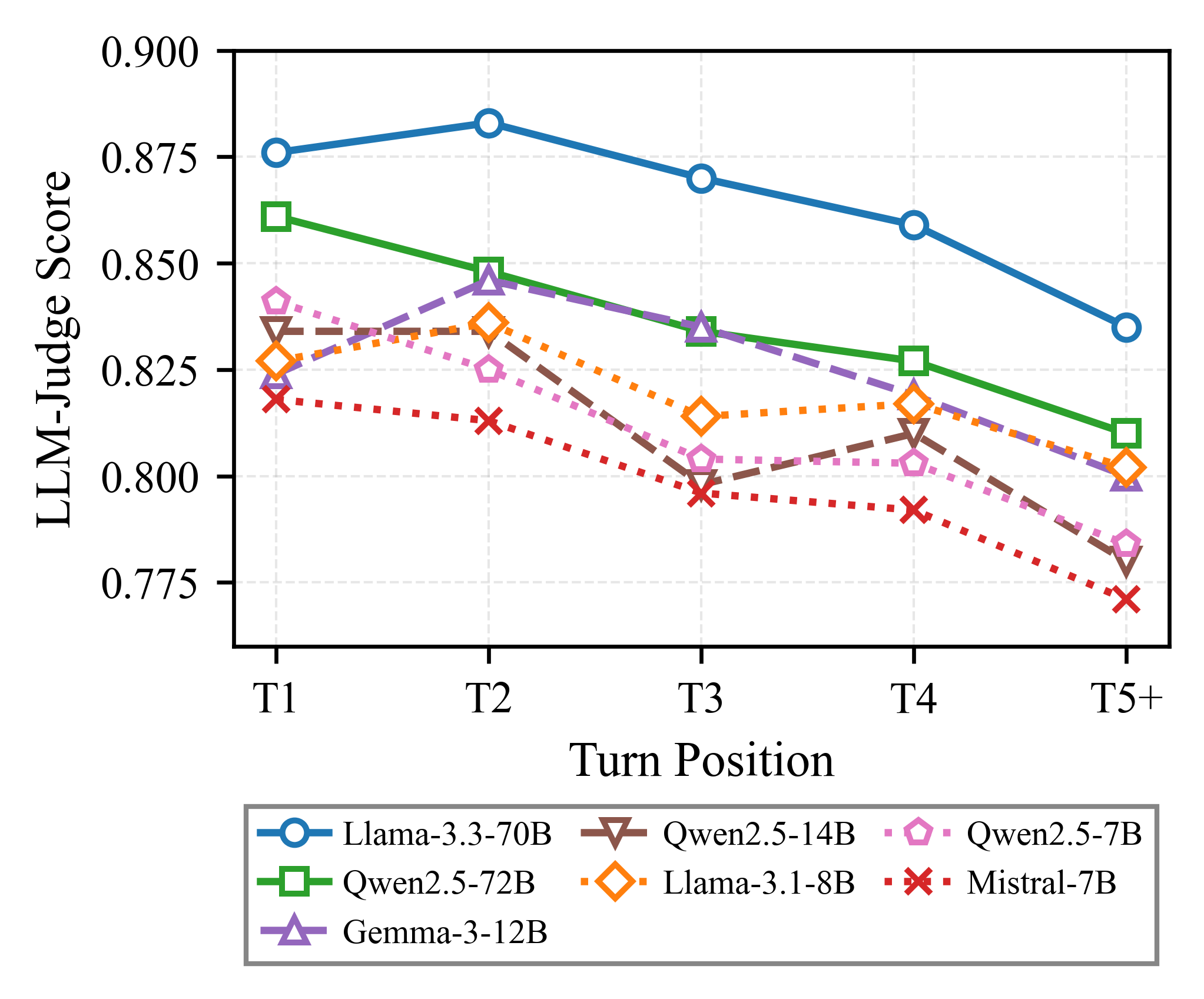}
    \caption{Generation quality by turn position across 7 models (Retrieved mode). Large models (solid) outperform small (dotted). All show significant decline ($p < 0.01$).}
    \label{fig:generation_turn}
\end{figure}

As shown in Figure~\ref{fig:generation_turn}, generation quality drops at later turns: in Retrieved mode, Llama-3.3-70B declines 4.1 points from T1 to T5+. In contrast, passage retrieval with History+Reasoning slightly improves across turns (\S\ref{sec:turn_position}), suggesting that accumulated context benefits retrievers more than generators.



\paragraph{Dimension analysis.} Across all models, Coherence scores highest (.895--.940), showing that LLMs produce fluent responses. Completeness scores lowest (.750--.846), indicating that covering all aspects of complex questions remains challenging. (See Appendix~\ref{sec:app_dimension} for details)


\section{Conclusion}


This work introduces the first benchmark combining multi-turn dialogue with reasoning-intensive retrieval. Our Decomposition and Verification framework generates fact grounded conversations through multi-level validation. Evaluation confirms conversational context is critical: History+Reasoning doubles baseline performance (.479 vs .236 nDCG@10). reasoning-specialized models outperform dense retrievers (DIVER .584 vs SFR .464), showing that multi-hop reasoning generalizes to conversation. However, implicit reasoning dependent on unstated domain knowledge remains challenging. These findings highlight current capabilities and limitations, directing future research in reasoning-guided conversational retrieval.


\section*{Limitations}

\noindent\textbf{(1) Language coverage.} The benchmark covers English only. Extending to other languages would require adapting the verification pipeline and collecting multilingual domain corpora.\\
\textbf{(2) Domain scope.} We cover eleven domains spanning scientific, technical, and social topics, but areas such as mathematics and commonsense reasoning remain unexplored and may exhibit different performance patterns.\\
\textbf{(3) Source dependency.} Our framework requires existing question-answer pairs with supporting documents. Domains lacking such structured resources would require additional annotation.\\

\FloatBarrier
\bibliography{custom}
\clearpage

\appendix

\section{Extended Related Work}
\label{sec:extended_related_work}

\paragraph{Conversational Information Retrieval.}
Conversational search benchmarks have established foundations for multi-turn retrieval evaluation. QReCC \citep{anantha2021open} provides 14K conversations emphasizing question rewriting to resolve coreferences and ellipses. TREC iKAT \citep{aliannejadi2024trec} adds personalization through user-specific knowledge bases over 116 million passages. MTRAG \citep{katsis2025mtrag} offers 110 human-generated conversations with challenges including unanswerable questions. The TREC Conversational Assistance Track (CAsT) \citep{dalton2020cast} introduced conversational passage retrieval with artifacts such as ellipsis and anaphora, driving advances in conversational dense retrieval \citep{yu2021convdr}. These benchmarks evaluate conversational dynamics but assume relevance depends on semantic similarity rather than reasoning.
Related work has also addressed language-specific challenges in retrieval and understanding, including aspect-based sentiment analysis for Arabic text \citep{abdelgwad2022arabic}.

\paragraph{Multi-turn RAG and Conversational QA.}
Additional benchmarks target multi-turn retrieval-augmented generation. CORAL \citep{yue2024coral} provides large-scale conversations from Wikipedia supporting retrieval, generation, and citation tasks. Doc2Dial \citep{feng2020doc2dial} creates goal-oriented dialogues grounded in government documents. OR-QuAC \citep{qu2020orquac} extends QuAC to open-retrieval settings.

Foundational conversational QA datasets include CoQA \citep{reddy2019coqa} with 127K questions featuring coreference and pragmatic reasoning, QuAC \citep{choi2018quac} with 14K information-seeking dialogues, and TopiOCQA \citep{adlakha2022topiocqa} introducing topic switching. These datasets evaluate context resolution but do not require reasoning-intensive retrieval.

\paragraph{Reasoning-Intensive Retrieval.}
BRIGHT \citep{su2025bright} revealed that retrievers achieving 59.0 nDCG@10 on standard benchmarks drop to 18.3 on reasoning-intensive queries where relevant documents share no surface terms with queries. Chain-of-thought augmentation improves performance by up to 12.2 points. However, BRIGHT evaluates only single-turn queries, leaving multi-turn reasoning-intensive retrieval unexplored.
Recent advances in reasoning-enhanced retrieval include reranking approaches that leverage LLM distillation \citep{abdallah2025dear} and zero-shot methods using answer scent signals \citep{abdallah2025asrank}.

\paragraph{RAG Evaluation.}
RAG benchmarks assess grounded generation quality. RGB \citep{chen2024benchmarking} tests noise robustness and information integration. RAGBench \citep{friel2024ragbench} provides fine-grained annotations. CRAG \citep{yang2024crag} covers diverse question types across domains. These primarily evaluate single-turn interactions.
Complementary efforts have developed unified evaluation platforms for retrieval, reranking, and RAG systems \citep{abdallah2025rerankarena}, along with comprehensive toolkits supporting end-to-end RAG pipelines \citep{abdallah2025rankify}. Domain-specific benchmarks have also emerged for specialized areas such as corporate sustainability reporting \citep{ali2025sustainableqa}.

\paragraph{Atomic Fact Verification.}
Our verification pipeline relates to factuality evaluation methods. FActScore \citep{min2023factscore} decomposes generations into atomic facts validated against Wikipedia. SAFE \citep{wei2024safe} extends this with search-augmented verification. SelfCheckGPT \citep{manakul2023selfcheckgpt} detects hallucinations through sampling consistency. Unlike these post-hoc methods, we use atomic fact extraction during construction to ensure each turn is grounded in source documents.

\paragraph{Multi-hop Reasoning.}
Multi-hop QA benchmarks require synthesizing information across documents. HotpotQA \citep{yang2018hotpotqa} provides 113K questions with sentence-level supporting facts. 2WikiMultiHopQA \citep{ho2020constructing} constructs 2-hop and 4-hop questions via shared entities. MuSiQue \citep{trivedi2022musique} addresses shortcut exploitation through bottom-up composition. These focus on comprehension given retrieved passages; BRIGHT and our work evaluate whether retrievers can identify relevant passages when relevance requires reasoning.

\paragraph{Positioning Our Work.}
Existing benchmarks address subsets of the challenges in complex information seeking. Conversational benchmarks (QReCC, CAsT, MTRAG, CORAL) evaluate multi-turn dynamics without reasoning-intensive retrieval. BRIGHT requires reasoning but operates single-turn. Multi-hop QA benchmarks (HotpotQA, MuSiQue) require reasoning but assume gold retrieval.

Our benchmark uniquely combines: (1) multi-turn conversation with natural dialogue flow, (2) reasoning-intensive retrieval where relevance requires inference, (3) fact-grounded responses verified against source documents, and (4) explicit retrieval reasoning extending BRIGHT's insights to conversational settings. This combination tests whether systems can maintain both conversational coherence and reasoning-based retrieval across extended interactions---essential for supporting complex real-world information needs.
\section{Detailed Dataset Statistics}
\label{app:detailed_stats}

\subsection{Turn Distribution}

Conversation lengths range from 3 to 12 turns, with the distribution shown in Table~\ref{tab:turn_dist}. Most conversations contain 3--4 turns, while complex topics requiring extensive multi-step reasoning extend to longer dialogues.

\begin{table}[htbp]
\centering
\small
\begin{tabular}{lrr}
\toprule
\textbf{Turns} & \textbf{Count} & \textbf{Percentage} \\
\midrule
3 & 296 & 41.9\% \\
4 & 195 & 27.6\% \\
5 & 108 & 15.3\% \\
6 & 50 & 7.1\% \\
7+ & 58 & 8.2\% \\
\bottomrule
\end{tabular}
\caption{Turn distribution across all 707 conversations (aggregated from 11 domains).}
\label{tab:turn_dist}
\end{table}

\subsection{Aspect Distribution}

Each conversation contains multiple distinct semantic aspects extracted from the original answer. Table~\ref{tab:aspect_dist} shows the distribution across conversations, demonstrating that most conversations contain 5--7 aspects, with variation based on answer complexity.

\begin{table}[htbp]
\centering

\small
\begin{tabular}{lrr}
\toprule
\textbf{Aspects} & \textbf{Count} & \textbf{Percentage} \\
\midrule
3 & 82 & 11.6\% \\
4 & 121 & 17.1\% \\
5 & 161 & 22.8\% \\
6 & 100 & 14.1\% \\
7 & 93 & 13.2\% \\
8+ & 150 & 21.2\% \\
\bottomrule
\end{tabular}
\caption{Distribution of semantic aspects per conversation across all 707 conversations (aggregated from 11 domains).}
\label{tab:aspect_dist}
\end{table}

\subsection{Document Diversity per Turn}

Supporting documents per turn range from single focused sources to synthesis across multiple documents. Table~\ref{tab:doc_dist} shows this natural distribution, where simpler aspects may be addressed with one document while complex reasoning requires multiple sources.

\begin{table}[htbp]
\centering

\small
\begin{tabular}{lrr}
\toprule
\textbf{Docs/Turn} & \textbf{Count} & \textbf{Percentage} \\
\midrule
1 & 1,667 & 56.1\% \\
2 & 527 & 17.7\% \\
3 & 312 & 10.5\% \\
4 & 248 & 8.3\% \\
5+ & 217 & 7.3\% \\
\bottomrule
\end{tabular}
\caption{Distribution of supporting documents per turn across all 2,971 turns (aggregated from 11 domains)}
\label{tab:doc_dist}
\end{table}


\section{Pipeline Illustration}
\label{appendix:pipeline}

This appendix illustrates our Decomposition-and-Verification framework through a complete example, tracing a single source query through all pipeline stages to produce a grounded multi-turn conversation.

\subsection{Source Material}

\paragraph{Original Query.}
\begin{quote}
\textit{``Why does evolution not make our life longer? Wouldn't evolution favour a long life?''}
\end{quote}

\paragraph{Reasoning Annotation.}
\begin{quote}
\textit{``The question requires understanding why natural selection does not optimize for longevity. Relevant documents must explain evolutionary theories of aging, including mutation accumulation, antagonistic pleiotropy, and life-history trade-offs.''}
\end{quote}

\subsection{Stage 1: Answer Decomposition}

The expert answer addresses multiple evolutionary mechanisms. Our pipeline identifies six distinct aspects, each suitable for a conversational turn.

\begin{table}[htbp]
\centering
\small

\vspace{0.3em}
\begin{tabular}{@{}clcc@{}}
\toprule
\textbf{\#} & \textbf{Aspect} & \textbf{Type} & \textbf{Cov.} \\
\midrule
1 & Post-reproductive selection & Mechanism & 0.85 \\
2 & Mutation accumulation & Mechanism & 0.90 \\
3 & Late-onset disease example & Example & 0.80 \\
4 & Antagonistic pleiotropy & Mechanism & 0.85 \\
5 & Disposable soma theory & Mechanism & 0.75 \\
6 & Extrinsic mortality & Implication & 0.80 \\
\bottomrule
\end{tabular}
\caption{Aspects identified from the source answer.}
\label{tab:aspects}
\end{table}

\subsection{Stage 2--3: Fact Extraction and Verification}
From each aspect, we extract atomic facts and verify them against source documents using semantic similarity. Facts without clear document support are discarded. Aspects with no verified facts are excluded from conversation generation.

Tables~\ref{tab:aspects}, \ref{tab:verification}, and \ref{tab:scoring} illustrate the pipeline progression. Table~\ref{tab:aspects} shows the six aspects decomposed from the source answer (Stage 1), along with a coverage score (Cov.) indicating how much of each aspect is topically addressed by source documents. Table~\ref{tab:verification} shows the atomic facts extracted from each aspect and their verification status (Stages 2--3).

\begin{table*}[t]
\centering
\small
\vspace{0.3em}
\begin{tabular}{@{}p{3.2cm}p{9.5cm}cc@{}}
\toprule
\textbf{Aspect} & \textbf{Extracted Facts} & \textbf{Verified} & \textbf{Rate} \\
\midrule
Post-reproductive selection & 
(1) Selection acts primarily on pre-reproductive traits; 
(2) Post-reproductive mutations face weak selection pressure & 2/2 & 100\% \\[0.8em]
Mutation accumulation & 
(1) Deleterious late-onset alleles accumulate over generations & 1/1 & 100\% \\[0.8em]
Late-onset disease example & 
(1) Huntington's manifests between ages 30--50; 
(2) Carriers reproduce before experiencing symptoms & 2/2 & 100\% \\[0.8em]
Antagonistic pleiotropy & 
(1) Certain alleles confer early-life benefits but late-life costs & 1/1 & 100\% \\[0.8em]
Disposable soma theory & 
(1) Organisms allocate resources between reproduction and repair; 
(2) High repair investment yields longer lifespan & 0/2 & 0\% \\[0.8em]
Extrinsic mortality & 
(1) High environmental mortality favors early reproduction & 1/1 & 100\% \\
\midrule
\textbf{Total} & & \textbf{7/9} & \textbf{78\%} \\
\bottomrule
\end{tabular}
\caption{Fact extraction and verification results. Each row shows the atomic facts extracted from an aspect and whether they were verified against source documents.}
\label{tab:verification}
\end{table*}

As shown in Table~\ref{tab:verification}, the disposable soma theory aspect yielded no verified facts and is therefore excluded from the generated conversation. This illustrates an important distinction: high coverage does not guarantee fact verification. The disposable soma theory has a coverage score of 0.75 in Table~\ref{tab:aspects}, meaning the source documents discuss this topic. However, its verification rate is 0\% because the two specific claims extracted from the gold answer could not be confirmed in those documents. The remaining five aspects form the basis for the five-turn dialogue.

Table~\ref{tab:scoring} demonstrates document scoring for Turn 4, where candidate documents are evaluated based on their support for the verified facts (Stage 4).

\subsection{Stage 4: Document Scoring}

Candidate documents are scored using:
\begin{equation*}
\text{score} = 0.5 S_{\text{s}} + 0.3 S_{\text{c}} + 0.15 S_{\text{l}} - 0.05 S_{\text{m}}
\end{equation*}
where $S_{\text{s}}$, $S_{\text{c}}$, $S_{\text{l}}$, $S_{\text{m}}$ denote support, completeness, clarity, and misleading scores respectively.

\begin{table}[htbp]
\centering
\small
\vspace{0.3em}
\setlength{\tabcolsep}{4pt}
\begin{tabular}{@{}lcccccc@{}}
\toprule
\textbf{Document} & $S_{\text{s}}$ & $S_{\text{c}}$ & $S_{\text{l}}$ & $S_{\text{m}}$ & \textbf{Score} & \textbf{Sel.} \\
\midrule
Aging\_theory\_12.txt & 8 & 7 & 7 & 0 & 7.15 & \checkmark \\
Evolution\_38.txt & 6 & 5 & 6 & 0 & 5.40 & \checkmark \\
Genetics\_91.txt & 4 & 3 & 5 & 0 & 3.65 & --- \\
\bottomrule
\end{tabular}
\caption{Document scoring example (Turn 4). Documents above threshold are selected.}
\label{tab:scoring}
\end{table}

\subsection{Generated Conversation}

The pipeline produces the following five-turn conversation from the five aspects with verified facts.

\vspace{0.3em}
\begin{tcolorbox}[colback=white, colframe=black!60, boxrule=0.4pt, arc=0pt, left=3pt, right=3pt, top=2pt, bottom=2pt, boxsep=0pt, before skip=0pt, after skip=-0.4pt]
\small
\textbf{Turn 1} \hfill \textit{Post-reproductive selection}

\textbf{Q:} In evolution, why doesn't natural selection care much about traits that show up after we've had kids?

\textbf{A:} Natural selection acts primarily on traits affecting survival and reproduction before the end of the reproductive period. Mutations with effects manifesting only after reproduction face substantially weaker selection pressure, since individuals have already transmitted their genes.
\end{tcolorbox}
\begin{tcolorbox}[colback=white, colframe=black!60, boxrule=0.4pt, arc=0pt, left=3pt, right=3pt, top=2pt, bottom=2pt, boxsep=0pt, before skip=-0.4pt, after skip=-0.4pt]
\small
\textbf{Turn 2} \hfill \textit{Mutation accumulation}

\textbf{Q:} Is that why some harmful genes stick around, just because they show up later in life?

\textbf{A:} Precisely. This is the basis of mutation accumulation theory. Deleterious alleles with late-onset effects are not efficiently removed by selection because their harmful consequences occur after reproduction. Over generations, such mutations accumulate in the population.
\end{tcolorbox}
\begin{tcolorbox}[colback=white, colframe=black!60, boxrule=0.4pt, arc=0pt, left=3pt, right=3pt, top=2pt, bottom=2pt, boxsep=0pt, before skip=-0.4pt, after skip=-0.4pt]
\small
\textbf{Turn 3} \hfill \textit{Late-onset example}

\textbf{Q:} Are there examples of genetic diseases that stick around because they only appear after people have children?

\textbf{A:} Huntington's disease is a classic example. Symptoms typically manifest between ages 30 and 50, often after individuals have had children. Because carriers reproduce before experiencing symptoms, selection against the causative allele is weak.
\end{tcolorbox}
\begin{tcolorbox}[colback=white, colframe=black!60, boxrule=0.4pt, arc=0pt, left=3pt, right=3pt, top=2pt, bottom=2pt, boxsep=0pt, before skip=-0.4pt, after skip=-0.4pt]
\small
\textbf{Turn 4} \hfill \textit{Antagonistic pleiotropy}

\textbf{Q:} What about genes that actually help us when we're young but cause problems when we're older?

\textbf{A:} This describes antagonistic pleiotropy, proposed by George Williams in 1957. Certain alleles confer fitness benefits early in life but produce deleterious effects later. Selection favours these alleles because early-life benefits outweigh late-life costs.
\end{tcolorbox}
\begin{tcolorbox}[colback=white, colframe=black!60, boxrule=0.4pt, arc=0pt, left=3pt, right=3pt, top=2pt, bottom=2pt, boxsep=0pt, before skip=-0.4pt, after skip=0pt]
\small
\textbf{Turn 5} \hfill \textit{Extrinsic mortality}

\textbf{Q:} Why wouldn't living longer be advantageous if there's a lot of danger in the environment?

\textbf{A:} When extrinsic mortality is high, few individuals survive to benefit from extended longevity. Under such conditions, selection favours early reproduction over investment in long-term survival, explaining why high-mortality species typically have shorter lifespans.
\end{tcolorbox}

\vspace{0.3em}

\subsection{Reasoning Requirements}

Each turn requires reasoning to connect the question to relevant documents. Table~\ref{tab:reasoning} shows the inference needed for each turn.

\begin{table*}[t]
\centering
\small
\vspace{0.3em}
\begin{tabular}{@{}c @{\hspace{1em}} p{4.0cm} @{\hspace{1em}} p{3.5cm} @{\hspace{1em}} p{5.5cm}@{}}
\toprule
\textbf{Turn} & \textbf{Question} & \textbf{Document Topic} & \textbf{Reasoning Required} \\
\midrule
1 & Traits appearing after having children & Post-reproductive selection dynamics & Recognize that timing of trait expression determines selection intensity \\[0.6em]
2 & Harmful genes persisting in populations & Late-acting mutation accumulation & Infer that weak selection permits mutation buildup over generations \\[0.6em]
3 & Diseases appearing after reproduction & Huntington's disease onset patterns & Identify specific disease as instance of the general pattern \\[0.6em]
4 & Genes helping early, harming later & Antagonistic pleiotropy & Map everyday description to formal evolutionary concept \\[0.6em]
5 & Longevity despite environmental danger & Extrinsic mortality effects & Connect environmental risk to selection on lifespan \\
\bottomrule
\end{tabular}
\caption{Reasoning required for retrieval. Finding relevant documents requires understanding the conceptual relationship between what the user asks and what documents discuss.}
\label{tab:reasoning}
\end{table*}

These reasoning requirements are inherent to the domain. A retrieval system must recognize, for example, that documents about ``antagonistic pleiotropy'' are relevant to a question about ``genes that help early but cause problems later.'' This requires understanding the conceptual relationship between the question and the document topic---a process that goes beyond surface-level matching.

\subsection{Retrieval Reasoning Annotation}

Each turn includes annotations specifying what information relevant documents should contain. Example for Turn 4:

\begin{lstlisting}[basicstyle=\ttfamily\small, frame=single, xleftmargin=2pt, xrightmargin=2pt]
{
  "target": "Genes with age-dependent
             opposing fitness effects",
  "relevance_signals": [
    "antagonistic pleiotropy",
    "early-life benefits with late-life costs",
    "Williams 1957 aging theory"
  ],
  "irrelevance_signals": [
    "neutral mutations",
    "single-effect alleles"
  ]
}
\end{lstlisting}

\section{Additional Retrieval Results}
\label{sec:app_retrieval}

\subsection{Statistical Details}
\label{sec:app_stats}

Table~\ref{tab:stats} presents detailed statistical comparisons referenced in \S\ref{sec:retrieval_results}.

\begin{table}[htbp]
\centering

\small
\begin{tabular}{l|cccc}
\toprule
\textbf{Comparison} & \textbf{Mean $\Delta$} & \textbf{SE} & \textbf{$t$} & \textbf{$p$} \\
\midrule
QR vs.\ Baseline & +0.086 & 0.0016 & 61.1 & $<$0.001 \\
Reas vs.\ Baseline & +0.157 & 0.0019 & 92.9 & $<$0.001 \\
Hist vs.\ Baseline & +0.204 & 0.0022 & 102.2 & $<$0.001 \\
H+R vs.\ Baseline & +0.243 & 0.0022 & 119.7 & $<$0.001 \\
H+R vs.\ History & +0.039 & 0.0013 & 30.2 & $<$0.001 \\
\bottomrule
\end{tabular}
\caption{Paired t-tests ($n=23,768$). Mean $\Delta$: macro-averaged nDCG@10 difference across domains. SE: std. error.}

\label{tab:stats}
\end{table}

\subsection{Additional Metrics}
\label{sec:app_metrics}

Table~\ref{tab:full_metrics} presents MAP@10, Recall@10, and MRR for History+Reasoning.

  \begin{table}[htbp]
  \centering
  
  \small
  \begin{tabular}{lcccc}
  \toprule
  \textbf{Retriever} & \textbf{nDCG@10} & \textbf{MAP@10} & \textbf{Recall@10} & \textbf{MRR} \\
  \midrule
  DIVER & .584 & .488 & .753 & .592 \\
  ReasonIR & .552 & .459 & .713 & .566 \\
  BM25 & .489 & .406 & .627 & .508 \\
  SFR & .464 & .374 & .627 & .470 \\
  Qwen & .461 & .368 & .633 & .464 \\
  BGE & .445 & .356 & .595 & .463 \\
  E5 & .429 & .350 & .573 & .437 \\
  Contriever & .409 & .325 & .552 & .428 \\
  \bottomrule
  \end{tabular}
  \caption{Full metrics for History+Reasoning (all @10, macro-averaged across 11 domains).}
  \label{tab:full_metrics}
  \end{table}

\subsection{Turn Position by Retriever}
\label{sec:app_turn}

Table~\ref{tab:turn_retriever} shows per-retriever turn position effects referenced in \S\ref{sec:turn_position}.

\begin{table}[htbp]
\centering

\small
\setlength{\tabcolsep}{3pt}
\begin{tabular}{l|ccccc}
\toprule
\textbf{Retriever} & \textbf{T2} & \textbf{T3} & \textbf{T4} & \textbf{T5+} & \textbf{Trend} \\
\midrule
\multicolumn{6}{c}{\textit{History+Reasoning}} \\
\midrule
DIVER & .629 & .638 & .641 & .632 & +2\% \\
ReasonIR & .593 & .593 & .615 & .600 & +4\% \\
BM25 & .580 & .574 & .587 & .572 & +1\% \\
Qwen & .503 & .488 & .529 & .528 & +5\% \\
\midrule
\multicolumn{6}{c}{\textit{Baseline}} \\
\midrule
DIVER & .339 & .327 & .320 & .333 & $-$6\% \\
ReasonIR & .244 & .240 & .234 & .237 & $-$4\% \\
BM25 & .192 & .170 & .151 & .162 & $-$21\% \\
Qwen & .256 & .250 & .252 & .267 & $-$2\% \\
\bottomrule
\end{tabular}
\caption{Turn position by retriever. Trend shows T2$\rightarrow$T4 change. With History+Reasoning, all retrievers maintain or improve at later turns. With Baseline, all retrievers degrade.}
\label{tab:turn_retriever}
\end{table}

\subsection{Complete Domain Results}
\label{sec:app_domain}

Table~\ref{tab:domain_full} presents the full domain-by-retriever matrix referenced in \S\ref{sec:domain}.

\begin{table*}[t]
\centering

\small
\setlength{\tabcolsep}{10pt}
\begin{tabular}{l|cccccccc}
\toprule
\textbf{Domain} & \textbf{BGE} & \textbf{BM25} & \textbf{Contr.} & \textbf{DIVER} & \textbf{E5} & \textbf{Qwen} & \textbf{ReasIR} & \textbf{SFR} \\
\midrule
Biology & .463 & .722 & .459 & \textbf{.732} & .589 & .636 & .636 & .554 \\
Drones & .423 & .300 & .366 & .491 & .296 & .340 & \textbf{.499} & .359 \\
Earth Sci. & .531 & .687 & .480 & \textbf{.694} & .595 & .545 & .621 & .558 \\
Economics & .419 & .526 & .402 & \textbf{.572} & .496 & .383 & .527 & .507 \\
Hardware & .356 & .345 & .280 & .433 & .336 & .362 & \textbf{.455} & .358 \\
Law & .373 & .337 & .376 & .490 & .282 & .416 & \textbf{.540} & .373 \\
Medical & .421 & .203 & .380 & \textbf{.488} & .187 & .392 & .400 & .352 \\
Politics & .463 & .406 & .459 & \textbf{.582} & .309 & .478 & .557 & .391 \\
Psychology & .520 & .649 & .478 & \textbf{.733} & .588 & .601 & .695 & .591 \\
Robotics & .381 & \textbf{.587} & .358 & .529 & .489 & .355 & .526 & .469 \\
Sust.\ Living & .541 & .612 & .458 & \textbf{.682} & .552 & .568 & .618 & .586 \\
\bottomrule
\end{tabular}
\caption{Domain-retriever matrix (nDCG@10, History+Reasoning). Bold indicates best retriever per domain.}
\label{tab:domain_full}
\end{table*}

\section{Additional Generation Results}
\label{sec:app_generation}

\subsection{Per-Domain Generation Quality by Mode}
\label{sec:app_gen_domain}

Table~\ref{tab:gen_oracle} presents generation quality for Oracle mode across all domains.

\begin{table*}[t]
\centering

\small
\setlength{\tabcolsep}{3pt}
\begin{tabular}{l|ccccccc}
\toprule
\textbf{Domain} & \textbf{Llama-70B} & \textbf{Qwen-72B} & \textbf{Gemma-12B} & \textbf{Qwen-14B} & \textbf{Llama-8B} & \textbf{Qwen-7B} & \textbf{Mistral-7B} \\
\midrule
Psychology & \textbf{.932} & .878 & .895 & .889 & .870 & .856 & .862 \\
Biology & \textbf{.925} & .882 & .891 & .867 & .865 & .860 & .855 \\
Medical Sci. & \textbf{.918} & .871 & .886 & .875 & .862 & .848 & .845 \\
Earth Science & .812 & \textbf{.824} & .778 & .752 & .795 & .788 & .762 \\
Hardware & \textbf{.922} & .869 & .880 & .862 & .851 & .849 & .840 \\
Robotics & \textbf{.905} & .873 & .865 & .868 & .845 & .852 & .822 \\
Drones & \textbf{.912} & .862 & .898 & .865 & .871 & .846 & .838 \\
Economics & \textbf{.928} & .895 & .902 & .882 & .873 & .871 & .865 \\
Law & \textbf{.915} & .872 & .881 & .870 & .856 & .842 & .844 \\
Politics & \textbf{.896} & .868 & .879 & .862 & .848 & .835 & .838 \\
Sust.\ Living & \textbf{.919} & .870 & .885 & .872 & .862 & .849 & .848 \\
\midrule
\textbf{Average} & \textbf{.920} & .876 & .890 & .874 & .862 & .849 & .849 \\
\bottomrule
\end{tabular}
\caption{Generation quality by domain (Oracle mode). LLM-Judge average score. Bold indicates best model per domain.}
\label{tab:gen_oracle}
\end{table*}

Table~\ref{tab:gen_noret} presents generation quality for NoRetrieval mode across all domains.

\begin{table*}[t]
\centering

\small
\setlength{\tabcolsep}{3pt}
\begin{tabular}{l|ccccccc}
\toprule
\textbf{Domain} & \textbf{Llama-70B} & \textbf{Qwen-72B} & \textbf{Gemma-12B} & \textbf{Qwen-14B} & \textbf{Llama-8B} & \textbf{Qwen-7B} & \textbf{Mistral-7B} \\
\midrule
Psychology & \textbf{.872} & .815 & .822 & .828 & .762 & .778 & .758 \\
Biology & \textbf{.868} & .832 & .810 & .765 & .758 & .812 & .742 \\
Medical Sci. & \textbf{.862} & .818 & .815 & .822 & .765 & .795 & .738 \\
Earth Science & .698 & \textbf{.725} & .658 & .612 & .682 & .688 & .645 \\
Hardware & \textbf{.865} & .820 & .812 & .805 & .752 & .792 & .745 \\
Robotics & \textbf{.842} & .822 & .792 & .815 & .745 & .798 & .722 \\
Drones & \textbf{.848} & .812 & .835 & .808 & .778 & .788 & .742 \\
Economics & \textbf{.872} & .848 & .832 & .825 & .782 & .818 & .778 \\
Law & \textbf{.855} & .822 & .815 & .818 & .762 & .785 & .752 \\
Politics & \textbf{.835} & .818 & .818 & .808 & .752 & .778 & .748 \\
Sust.\ Living & \textbf{.862} & .822 & .822 & .818 & .772 & .795 & .758 \\
\midrule
\textbf{Average} & \textbf{.859} & .837 & .824 & .824 & .789 & .803 & .779 \\
\bottomrule
\end{tabular}
\caption{Generation quality by domain (NoRetrieval mode). LLM-Judge average score. Bold indicates best model per domain. Faithfulness excluded.}
\label{tab:gen_noret}
\end{table*}

\subsection{Dimension Breakdown}
\label{sec:app_dimension}

Table~\ref{tab:gen_dimension} presents dimension-level scores for all models using Retrieved mode.

\begin{table}[t]
\centering

\small
\setlength{\tabcolsep}{3pt}
\begin{tabular}{l|cccccc}
\toprule
\textbf{Model} & \textbf{Corr} & \textbf{Comp} & \textbf{Rel} & \textbf{Coh} & \textbf{Faith} & \textbf{Avg} \\
\midrule
Llama-3.3-70B & .870 & .846 & .909 & .940 & .875 & \textbf{.888} \\
Qwen2.5-72B & .848 & .781 & .896 & .929 & .846 & .860 \\
Gemma-3-12B & .840 & .784 & .879 & .916 & .849 & .854 \\
Qwen2.5-14B & .840 & .776 & .888 & .920 & .838 & .852 \\
Llama-3.1-8B & .816 & .787 & .864 & .911 & .823 & .840 \\
Qwen2.5-7B & .816 & .760 & .868 & .912 & .820 & .835 \\
Mistral-7B & .799 & .750 & .847 & .895 & .806 & .819 \\
\midrule
\textbf{Average} & .833 & .783 & .879 & .918 & .837 & .850 \\
\bottomrule
\end{tabular}
\caption{Generation quality by dimension (Retrieved mode). Corr=Correctness, Comp=Completeness, Rel=Relevance, Coh=Coherence, Faith=Faithfulness. Scores macro-averaged across 11 domains.}
\label{tab:gen_dimension}
\end{table}

Table~\ref{tab:gen_dim_domain} presents dimension scores by domain for Llama-3.3-70B.

\begin{table}[t]
\centering

\small
\setlength{\tabcolsep}{3pt}
\begin{tabular}{l|ccccc}
\toprule
\textbf{Domain} & \textbf{Corr} & \textbf{Comp} & \textbf{Rel} & \textbf{Coh} & \textbf{Faith} \\
\midrule
Psychology & .885 & .862 & .918 & .948 & .888 \\
Biology & .882 & .858 & .915 & .945 & .885 \\
Medical Sci. & .872 & .848 & .905 & .942 & .872 \\
Earth Science & .752 & .728 & .795 & .885 & .778 \\
Hardware & .875 & .852 & .908 & .942 & .878 \\
Robotics & .855 & .832 & .888 & .935 & .865 \\
Drones & .858 & .835 & .892 & .938 & .868 \\
Economics & .880 & .858 & .912 & .945 & .882 \\
Law & .865 & .842 & .898 & .938 & .872 \\
Politics & .845 & .822 & .878 & .932 & .858 \\
Sust.\ Living & .872 & .850 & .905 & .940 & .875 \\
\midrule
\textbf{Average} & .870 & .846 & .909 & .940 & .875 \\
\bottomrule
\end{tabular}
\caption{Dimension scores by domain (Llama-3.3-70B, Retrieved). Corr=Correctness, Comp=Completeness, Rel=Relevance, Coh=Coherence, Faith=Faithfulness.}
\label{tab:gen_dim_domain}
\end{table}

\subsection{Lexical Metrics}
\label{sec:app_lexical}

Table~\ref{tab:lexical} presents lexical evaluation metrics for Retrieved mode.

\begin{table}[t]
\centering

\small
\setlength{\tabcolsep}{4pt}
\begin{tabular}{l|ccc}
\toprule
\textbf{Model} & \textbf{ROUGE-L} & \textbf{METEOR} & \textbf{BERT-F1} \\
\midrule
Llama-3.3-70B & .274 & \textbf{.345} & .365 \\
Qwen2.5-72B & \textbf{.283} & .264 & \textbf{.381} \\
Gemma-3-12B & .258 & .270 & .351 \\
Qwen2.5-14B & .283 & .269 & .375 \\
Llama-3.1-8B & .276 & .326 & .355 \\
Qwen2.5-7B & .275 & .271 & .367 \\
Mistral-7B & .272 & .286 & .352 \\
\bottomrule
\end{tabular}
\caption{Lexical metrics (Retrieved mode). Macro-averaged across 11 domains. Bold indicates best per metric.}
\label{tab:lexical}
\end{table}

Lexical metrics show lower absolute values than LLM-Judge scores due to the open-ended nature of conversational answers. METEOR correlates most strongly with LLM-Judge ($r$ = .72), while ROUGE-L shows weaker correlation ($r$ = .54).

\subsection{Model Size Comparison}
\label{sec:app_size}

Table~\ref{tab:model_size} compares models within the same family across all three modes.

\begin{table}[t]
\centering

\small
\setlength{\tabcolsep}{3pt}
\begin{tabular}{l|cccc}
\toprule
\textbf{Model} & \textbf{Oracle} & \textbf{Retr.} & \textbf{No.Retr.} & \textbf{RAG$\Delta$} \\
\midrule
\multicolumn{5}{l}{\textit{Qwen2.5 family}} \\
Qwen2.5-72B & .876 & .860 & .837 & +2.3 \\
Qwen2.5-14B & .874 & .852 & .824 & +2.8 \\
Qwen2.5-7B & .849 & .835 & .803 & +3.2 \\
\midrule
\multicolumn{5}{l}{\textit{Llama family}} \\
Llama-3.3-70B & .920 & .888 & .859 & +2.9 \\
Llama-3.1-8B & .862 & .840 & .789 & +5.1 \\
\bottomrule
\end{tabular}
\caption{Model size comparison by family. RAG$\Delta$ = Retrieved $-$ NoRetrieval ($\times$100). Scores macro-averaged across 11 domains.}
\label{tab:model_size}
\end{table}

Within Qwen2.5, scaling from 7B to 72B improves Retrieved scores by 2.5 points. Llama shows larger scaling effects: 70B outperforms 8B by 4.8 points. The RAG benefit decreases with model size, confirming that larger models' parametric knowledge partially substitutes for retrieval.

\FloatBarrier

\onecolumn
\section{Pipeline Prompts}
\label{appendix:prompts}

We use GPT-4.1~\citep{openai2025gpt41} for all conversation generation steps. This appendix provides the complete prompts used in our Decomposition-and-Verification framework.
\\\\\\\\\\
\begin{figure*}[htbp]
\centering

\begin{tcolorbox}[
  colback=gray!5,
  colframe=gray!75,
  title=Prompt for Document-Answer Alignment Validation,
  width=\textwidth
]
\small\ttfamily
You are evaluating whether documents can support an answer.

\textbf{GOLD ANSWER:}\\
\{gold\_answer\}

\textbf{DOCUMENTS:}\\
\{documents\}

\textbf{IMPORTANT:}
\begin{itemize}
  \item Documents don't need ALL details
  \item Core facts are enough
  \item Related information counts as support
\end{itemize}

\textbf{TASK:}
\begin{enumerate}
  \item Identify KEY CLAIMS (focus on main ideas, not minor details, max 3-5 claims)
  \item Check if documents contain supporting information
  \item Return coverage percentage
\end{enumerate}

\textbf{Return JSON:}\\
\{\\
~~"key\_claims": ["main claim 1", "main claim 2", ...],\\
~~"supported\_claims": ["claim1", ...],\\
~~"unsupported\_claims": ["claim2", ...],\\
~~"coverage\_percentage": 0.0-1.0,\\
~~"is\_sufficient": boolean\\
\}
\end{tcolorbox}
\caption{Prompt for validating document-answer alignment.}
\label{fig:prompt_alignment}
\end{figure*}

\begin{figure*}[!t]
\centering

\begin{tcolorbox}[
  colback=gray!5,
  colframe=gray!75,
  title=Prompt for Granular Aspect Extraction,
  width=\textwidth
]
\small\ttfamily
Extract \{num\_aspects\} distinct, GRANULAR aspects from the gold answer.

\textbf{QUERY:} \{query\}

\textbf{REASONING} (why this answer is relevant):\\
\{reasoning\}

\textbf{GOLD ANSWER:}\\
\{gold\_answer\}

\textbf{ALREADY EXTRACTED ASPECTS} (do NOT duplicate these):\\
\{existing\_aspects\}

\textbf{WHAT IS A GRANULAR ASPECT?}\\
An aspect does NOT need to be a broad topic. It can be a specific detail, a single step in a process, or a distinct implication.
\begin{itemize}
  \item Broad (AVOID): "How Photosynthesis Works" (Too big, covers everything)
  \item Granular (GOOD): "Role of Chlorophyll in Light Absorption"
  \item Granular (GOOD): "The Calvin Cycle's Carbon Fixation Step"
\end{itemize}

\textbf{ASPECT REQUIREMENTS:}
\begin{enumerate}
  \item SPECIFICITY: Drill down into details. A single sentence with a verified fact can be an aspect.
  \item VERBATIM EXCERPT: Copy exact text from gold answer.
  \item SUBSTANTIVE: Must contain facts, mechanisms, or examples.
  \item DISTINCT ANGLE: If a topic is already covered, look for a specific implication, limitation, or counter-example.
\end{enumerate}

\textbf{ASPECT TYPES:}\\
"detail" | "step" | "implication" | "distinction" | "definition" | "mechanism" | "example" | "comparison" | "history" | "application"

\textbf{Return JSON:}\\
\{\\
~~"aspects": [\\
~~~~\{\\
~~~~~~"aspect\_name": "Specific Name (3-6 words)",\\
~~~~~~"aspect\_type": "detail|step|implication|...",\\
~~~~~~"excerpt": "Exact verbatim text from gold answer",\\
~~~~~~"distinct\_from\_existing": "How this differs from existing aspects"\\
~~~~\}\\
~~],\\
~~"extraction\_notes": "Notes on finding distinct details"\\
\}
\end{tcolorbox}
\caption{Prompt for extracting granular aspects from gold answers.}
\label{fig:prompt_aspects}
\end{figure*}

\begin{figure*}[!t]
\centering

\begin{tcolorbox}[
  colback=gray!5,
  colframe=gray!75,
  title=Prompt for Atomic Fact Extraction and Verification,
  width=\textwidth
]
\small\ttfamily
Extract key facts from the EXCERPT, then verify each against the DOCUMENTS.

\textbf{ASPECT EXCERPT:}\\
\{aspect\_excerpt\}

\textbf{DOCUMENTS:}\\
\{documents\}

\textbf{TASK:}
\begin{enumerate}
  \item Extract \textbf{1 to 5} key facts from the excerpt (depending on length).
  \begin{itemize}
    \item If the excerpt is a single detail, extract just \textbf{1 fact}.
    \item Do NOT invent facts to fill a quota.
  \end{itemize}
  \item For each fact, check if it's supported by the documents.
  \item Return ONLY facts that are supported.
\end{enumerate}

\textbf{A fact is "supported" if:}
\begin{itemize}
  \item Explicitly stated in documents, OR
  \item Clearly implied by combining document information.
\end{itemize}

\textbf{Return JSON:}\\
\{\\
~~"extracted\_facts": [\\
~~~~\{\\
~~~~~~"fact": "the extracted fact (5-15 words)",\\
~~~~~~"is\_supported": boolean,\\
~~~~~~"supporting\_doc\_id": "doc\_X" or null,\\
~~~~~~"reason": "brief reason"\\
~~~~\}\\
~~],\\
~~"supported\_facts": ["fact1", "fact2", ...],\\
~~"unsupported\_facts": ["fact3", ...],\\
~~"summary": \{"total\_extracted": integer, "supported\_count": integer\}\\
\}
\end{tcolorbox}
\caption{Prompt for extracting and verifying atomic facts against source documents.}
\label{fig:prompt_facts}
\end{figure*}

\begin{figure*}[!t]
\centering

\begin{tcolorbox}[
  colback=gray!5,
  colframe=gray!75,
  title=Prompt for Aspect Overlap Detection,
  width=\textwidth
]
\small\ttfamily
Does the NEW ASPECT cover content already covered by EXISTING ASPECTS?

\textbf{NEW ASPECT:} \{aspect\_name\} (\{aspect\_type\})\\
\textbf{CONTENT:} \{excerpt\}

\textbf{EXISTING ASPECTS:}\\
\{existing\_aspects\_text\}

\textbf{WHAT COUNTS AS OVERLAP?}\\
Two aspects OVERLAP if they:
\begin{itemize}
  \item Make the EXACT SAME factual claims.
  \item Explain the EXACT SAME step of a process.
  \item Are merely rephrasing the same information.
\end{itemize}

\textbf{WHAT IS NOT OVERLAP? (DISTINCT CONTENT):}
\begin{itemize}
  \item Different steps of the same mechanism (e.g., "Step 1: Input" vs "Step 2: Processing")
  \item Different examples of the same concept
  \item Specific details vs General definitions
  \item Implications vs Mechanisms
  \item "What it is" vs "How it works" vs "Why it matters"
\end{itemize}

\textbf{DECISION:} Does the new aspect contain specific details, steps, or implications NOT in the existing aspects?

\textbf{Return JSON:}\\
\{\\
~~"has\_overlap": boolean,\\
~~"overlaps\_with": "name of overlapping aspect" or null,\\
~~"overlap\_type": "same\_claims" | "same\_examples" | "same\_step" | "no\_overlap",\\
~~"reasoning": "one sentence explanation"\\
\}
\end{tcolorbox}
\caption{Prompt for detecting content overlap between aspects.}
\label{fig:prompt_overlap}
\end{figure*}

\begin{figure*}[!t]
\centering

\begin{tcolorbox}[
  colback=gray!5,
  colframe=gray!75,
  title=Prompt for Aspect Ordering by Logical Progression,
  width=\textwidth
]
\small\ttfamily\raggedright

Order these topic aspects for natural conversation progression.

\textbf{CONVERSATION STRATEGY:} \{conversation\_strategy\}

\textbf{ASPECTS} (format: [type] name):\\
\{aspect\_summaries\}

\textbf{ORDERING PRINCIPLES:}
\begin{enumerate}[leftmargin=*, nosep]
  \item Foundational/definitional aspects first (what something IS)
  \item Then mechanisms/processes (how it WORKS)
  \item Then historical context or causes (WHY/WHEN)
  \item Then specific examples/cases (INSTANCES)
  \item End with implications/modern relevance (SO WHAT)
\end{enumerate}

\textbf{EXAMPLE GOOD ORDER:}
\begin{itemize}[leftmargin=*, nosep]
  \item[1.] [factual] "Definition of nation vs state"\\
        $\rightarrow$ first (foundational)
  \item[2.] [explanatory] "Distinction between political/cultural"\\
        $\rightarrow$ second
  \item[3.] [historical] "Colonial impact on borders"\\
        $\rightarrow$ third
  \item[4.] [factual] "Stateless nations examples"\\
        $\rightarrow$ fourth
  \item[5.] [explanatory] "Modern nation-state challenges"\\
        $\rightarrow$ last (implications)
\end{itemize}

\textbf{Return JSON:}\\
\{\\
~~"ordered\_indices": [0, 2, 1, 3],\\
~~"reasoning": "Brief explanation of ordering logic"\\
\}
\end{tcolorbox}
\caption{Prompt for reordering aspects to ensure natural conversational flow.}
\label{fig:prompt_ordering}
\end{figure*}

\begin{figure*}[!t]
\centering

\begin{tcolorbox}[
  colback=gray!5,
  colframe=gray!75,
  title=Prompt for Sub-question Generation,
  width=\textwidth
]
\small\ttfamily
Generate a focused sub-question for this aspect.

\textbf{QUERY:} \{query\}

\textbf{OVERALL REASONING:} \{overall\_reasoning\}

\textbf{ASPECT:} \{aspect\_name\}\\
\textbf{TYPE:} \{aspect\_type\}

\textbf{KEY FACTS:}\\
\{semantic\_facts\}

\textbf{PREVIOUS SUB-QUESTIONS:}\\
\{previous\_subquestions\}

\textbf{Generate a sub-question that:}
\begin{enumerate}
  \item Aligns with the overall reasoning and query intent
  \item Specifically targets this aspect
  \item Is distinct from previous sub-questions
  \item Can be answered using the facts listed
\end{enumerate}

\textbf{Return JSON:}\\
\{\\
~~"sub\_question": "The focused sub-question?",\\
~~"confidence": 0.0-1.0,\\
~~"reasoning": "Why this targets this aspect AND aligns with overall reasoning"\\
\}
\end{tcolorbox}
\caption{Prompt for generating focused sub-questions targeting specific aspects.}
\label{fig:prompt_subquestion}
\end{figure*}

\begin{figure*}[!t]
\centering

\begin{tcolorbox}[
  colback=gray!5,
  colframe=gray!75,
  title=Prompt for Document Scoring,
  width=\textwidth
]
\small\ttfamily
Score how well each document helps answer the sub-question.

\textbf{SUB-QUESTION:} \{sub\_question\}

\textbf{RETRIEVAL REASONING:} \{retrieval\_reasoning\}

\textbf{KEY FACTS TO COVER:}\\
\{semantic\_facts\}

\textbf{CANDIDATE DOCUMENTS:}\\
\{candidate\_docs\}

\textbf{SCORING CRITERIA} (each 0-10):
\begin{itemize}
  \item \textbf{support\_score}: How well does the document support answering the sub-question?
  \item \textbf{completeness\_score}: Does the document cover the key facts listed above?
  \item \textbf{clarity\_score}: Is the information clearly presented and easy to extract?
  \item \textbf{misleading\_score}: Does the document contain potentially misleading information? (PENALTY)
\end{itemize}

\textbf{Final score formula:}\\
$\text{score} = 0.5 \times S_s + 0.3 \times S_c + 0.15 \times S_l - 0.05 \times S_m$

\textbf{Return JSON:}\\
\{\\
~~"document\_scores": [\\
~~~~\{\\
~~~~~~"doc\_id": "doc\_0",\\
~~~~~~"support\_score": 8,\\
~~~~~~"completeness\_score": 7,\\
~~~~~~"clarity\_score": 9,\\
~~~~~~"misleading\_score": 0,\\
~~~~~~"final\_score": 7.85,\\
~~~~~~"reasoning": "Brief justification"\\
~~~~\}\\
~~]\\
\}
\end{tcolorbox}
\caption{Prompt for scoring candidate documents based on multiple criteria.}
\label{fig:prompt_scoring}
\end{figure*}

\begin{figure*}[!t]
\centering

\begin{tcolorbox}[
  colback=gray!5,
  colframe=gray!75,
  title=Prompt for Turn 1 Query Generation,
  width=\textwidth
]
\small\ttfamily
Transform the technical sub-question into a natural opening query for a conversation.

\textbf{ORIGINAL QUERY TOPIC:} \{original\_query\}

\textbf{TECHNICAL SUB-QUESTION:} \{sub\_question\}

\textbf{REQUIREMENTS:}
\begin{enumerate}
  \item This is Turn 1---there is NO conversation history yet.
  \item The question must stand alone and introduce the topic naturally.
  \item Keep the same technical content but make it sound like a curious person asking.
  \item Briefly mention the topic area so the question has context.
\end{enumerate}

\textbf{GOOD EXAMPLES:}
\begin{itemize}
  \item Technical: "What is the mechanism of phototaxis in insects?"
  \item Natural: "I've noticed bugs always fly toward lights at night. What's actually pulling them there?"
  \item Technical: "How does the nasal cycle regulate airflow?"
  \item Natural: "In smell perception, are there basic smells that combine like RGB colors?"
\end{itemize}

\textbf{Return JSON:}\\
\{\\
~~"conversational\_query": "The natural opening question WITH topic intro",\\
~~"kept\_technical\_content": true/false,\\
~~"natural\_language\_used": true/false\\
\}
\end{tcolorbox}
\caption{Prompt for generating the first turn query with topic introduction.}
\label{fig:prompt_turn1}
\end{figure*}

\begin{figure*}[!t]
\centering

\begin{tcolorbox}[
  colback=gray!5,
  colframe=gray!75,
  title=Prompt for Follow-up Query Generation,
  width=\textwidth
]
\small\ttfamily
Transform this into a natural follow-up question.

\textbf{TOPIC:} \{original\_query\}

\textbf{RECENT CONVERSATION:}\\
\{history\}

\textbf{TECHNICAL QUESTION TO ASK:} \{sub\_question\}

\textbf{PREVIOUS OPENERS USED:} \{previous\_starters\}

\textbf{CORE REQUIREMENTS:}
\begin{enumerate}
  \item Keep the SAME content/intent as the technical question
  \item Sound like a real curious person, not a student or interviewer
  \item VARY your opener---never repeat a starting word/phrase from "previous openers"
  \item VARY your structure---don't use the same question pattern as recent turns
\end{enumerate}

\textbf{NATURAL CONVERSATION PATTERNS:}
\begin{itemize}
  \item \textbf{Direct:} "What makes...", "How does...", "Why do..."
  \item \textbf{Curious:} "I wonder if...", "What about...", "How come..."
  \item \textbf{Confirming:} "Does that mean...", "Is that why...", "Would that..."
  \item \textbf{Probing:} "But what if...", "Even when...", "What happens if..."
  \item \textbf{Connecting:} "And does that...", "Then how...", "Which would mean..."
\end{itemize}

\textbf{LANGUAGE RULES:}
\begin{itemize}
  \item Use simple, everyday words; keep under 20 words
  \item Use pronouns (it, that, this, they) to connect to previous content
  \item Use contractions naturally (don't, isn't, wouldn't)
  \item Do NOT use academic language or filler transitions
\end{itemize}

\textbf{Return JSON:}\\
\{\\
~~"conversational\_query": "The natural follow-up question",\\
~~"transition\_type": "A/B/C/D/E",\\
~~"uses\_natural\_language": true/false,\\
~~"references\_previous\_content": true/false\\
\}
\end{tcolorbox}
\caption{Prompt for generating natural follow-up queries in multi-turn conversations.}
\label{fig:prompt_followup}
\end{figure*}

\begin{figure*}[!t]
\centering

\begin{tcolorbox}[
  colback=gray!5,
  colframe=gray!75,
  title=Prompt for Grounded Answer Generation,
  width=\textwidth
]
\small\ttfamily
Answer this question naturally using the information provided.

\textbf{CONVERSATION SO FAR:}\\
\{history\}

\textbf{CURRENT QUESTION:}\\
\{query\}

\textbf{AVAILABLE INFORMATION:}\\
\{documents\}

\textbf{HOW TO ANSWER:}
\begin{enumerate}
  \item Use ONLY information from the provided text above
  \item Speak naturally---explain as if talking to a curious friend
  \item Build on what was discussed earlier in the conversation
  \item Focus on WHAT YOU CAN EXPLAIN, not what you can't
  \item If the full answer isn't available, explain what IS known about the topic
  \item Keep answer conversational: 2-4 sentences, clear and direct
\end{enumerate}

\textbf{IMPORTANT RULES:}
\begin{itemize}
  \item[$\checkmark$] State facts directly and naturally
  \item[$\checkmark$] Connect to previous conversation smoothly
  \item[$\checkmark$] Explain using everyday language
  \item[$\times$] Do NOT mention "documents", "sources", "the text"
  \item[$\times$] Do NOT end with "However, [gaps in information]"
  \item[$\times$] Do NOT add invented examples, names, or details
\end{itemize}

\textbf{Return JSON:}\\
\{\\
~~"answer": "Your natural, conversational answer",\\
~~"uses\_natural\_language": true/false,\\
~~"avoids\_meta\_references": true/false,\\
~~"focuses\_on\_available\_info": true/false\\
\}
\end{tcolorbox}
\caption{Prompt for generating grounded answers from retrieved documents.}
\label{fig:prompt_answer}
\end{figure*}

\begin{figure*}[!t]
\centering

\begin{tcolorbox}[
  colback=gray!5,
  colframe=gray!75,
  title=Prompt for Turn Diversity Validation,
  width=\textwidth
]
\small\ttfamily
Does this new answer add value to the conversation?

\textbf{RECENT CONVERSATION:}\\
\{previous\_content\}

\textbf{NEW ANSWER:}\\
\{new\_answer\}

\textbf{EVALUATION CRITERIA:}

$\checkmark$ \textbf{ACCEPT} if answer does ANY of these:
\begin{enumerate}
  \item Introduces NEW factual information not previously stated
  \item Drills deeper into a specific detail (e.g., mechanism $\rightarrow$ sub-mechanism)
  \item Provides a concrete example of something described generally before
  \item Explains a consequence/implication not yet discussed
  \item Answers a different aspect of the same topic
\end{enumerate}

$\times$ \textbf{REJECT} only if answer does BOTH:
\begin{enumerate}
  \item Contains the SAME factual claims as previous turns (not just same topic)
  \item Uses similar phrasing/wording to express those facts
\end{enumerate}

\textbf{IMPORTANT DISTINCTIONS:}
\begin{itemize}
  \item Same TOPIC but different ANGLE = ACCEPT
  \item Same FACTS with different WORDS = REJECT
  \item General $\rightarrow$ Specific = ACCEPT
\end{itemize}

\textbf{Return JSON:}\\
\{\\
~~"adds\_value": boolean,\\
~~"value\_type": "new\_facts" | "deepening" | "example" | "implication" | "different\_angle" | "repetitive",\\
~~"reason": "brief explanation"\\
\}
\end{tcolorbox}
\caption{Prompt for validating that new turns add value and avoid repetition.}
\label{fig:prompt_diversity}
\end{figure*}

\begin{figure*}[!t]
\centering

\begin{tcolorbox}[
  colback=gray!5,
  colframe=gray!75,
  title=Prompt for Aspect Suitability Classification,
  width=\textwidth
]
\small\ttfamily
Should this aspect become a conversation sub-question?

\textbf{ASPECT:} \{aspect\_name\}\\
\textbf{TYPE:} \{aspect\_type\}\\
\textbf{EXCERPT:} \{excerpt\}

\textbf{CLASSIFICATION RULES:}
\begin{enumerate}
  \item \textbf{"substantive" $\rightarrow$ YES}: Contains facts, mechanisms, explanations, examples, or concrete information.
  \begin{itemize}
    \item NOTE: Short excerpts ARE ACCEPTABLE if they contain a clear fact.
  \end{itemize}
  \item \textbf{"meta" $\rightarrow$ NO}: Contains disclaimers, advice to "consult professional", caveats, or meta-commentary.
  \item \textbf{"insufficient" $\rightarrow$ NO}: Content is meaningless or purely structural (e.g., "Here is a list:").
\end{enumerate}

\textbf{Return JSON:}\\
\{\\
~~"should\_generate": boolean,\\
~~"reason": "explanation",\\
~~"aspect\_category": "substantive" | "meta" | "insufficient"\\
\}
\end{tcolorbox}
\caption{Prompt for filtering aspects based on suitability for conversation generation.}
\label{fig:prompt_suitability}
\end{figure*}

\clearpage
\twocolumn
\section{Automatic Quality Validation}
\label{appendix:automatic_validation}

\subsection{Overview}
\label{appendix:overview}

To complement the manual evaluation conducted on 200 randomly sampled conversations (Section~\ref{sec:human_evaluation}), we employed GPT-4o to automatically assess all conversations in the benchmark. This large-scale validation serves two purposes: (1) confirming that the quality observed in the manual sample generalizes to the full dataset, and (2) providing detailed turn-level linguistic analysis that would be impractical to obtain manually.

The automatic evaluation operates at two levels. At the conversation level, we assess the same four quality dimensions used in manual evaluation: naturalness, turn coherence, question quality, and groundedness. At the turn level, we analyze how each question depends on prior conversation history (dependency type) and what type of information each question seeks (question pattern).

\subsection{Conversation-Level Quality Assessment}
\label{appendix:conversation_quality}

We evaluated all 707 conversations on four quality dimensions using separate focused prompts for each dimension. Each dimension receives a score from 1 to 5, matching the Likert scale used in manual evaluation.

\textbf{Naturalness} measures whether the conversation sounds like natural human speech rather than robotic or overly formal language. The prompt checks for casual word choices, natural conversation starters, and appropriate sentence style.

\textbf{Turn Coherence} evaluates whether conversation turns connect logically to each other. The prompt examines clear references (pronouns pointing to previously mentioned content), logical connections between questions and preceding answers, and absence of abrupt topic jumps.

\textbf{Question Quality} assesses whether questions cover different useful aspects of the topic without repetition. The prompt checks that each question asks about something distinct, questions are specific enough to elicit clear answers, and together the questions provide good coverage of the topic.

\textbf{Groundedness} determines whether answers are supported by the source documents. The prompt verifies that claims match document content, no information is fabricated, and no contradictions exist between answers and sources.

Table~\ref{tab:llm_human_comparison} presents the comparison between LLM and human scores across all four dimensions.


\begin{table}[t]
\centering
\small
\setlength{\tabcolsep}{0pt} 
\begin{tabular*}{\columnwidth}{@{\extracolsep{\fill}}lcccc}
\toprule
\textbf{Dim.} & \textbf{LLM} & \textbf{$\sigma$} & \textbf{Hum.} & \textbf{$\Delta$} \\
\midrule
Naturalness & 4.34 & 0.58 & 4.20 & +0.14 \\
Turn Coherence & 4.97 & 0.18 & 4.62 & +0.35 \\
Quest. Quality & 4.42 & 0.54 & 4.21 & +0.21 \\
Groundedness & 4.73 & 0.53 & 4.45 & +0.28 \\
\bottomrule
\end{tabular*}
\caption{Comparison of automatic (LLM) and Human scores. \textbf{Dim.}: Quality Dimension; \textbf{LLM}: GPT-4o mean; \textbf{$\sigma$}: Standard deviation; \textbf{Hum.}: Human mean; \textbf{$\Delta$}: Difference (LLM $-$ Human). All differences are within the acceptable 0.5 threshold.}
\label{tab:llm_human_comparison}
\end{table}

The results demonstrate strong alignment between automatic and manual evaluation. All four dimensions show LLM scores within 0.35 points of human scores, well below the 0.5 threshold we consider acceptable. The consistently positive differences indicate that LLM evaluation tends to be slightly more lenient than human evaluation, a pattern commonly observed in LLM-as-judge settings. The low standard deviations, particularly for turn coherence (0.18), reflect high consistency in the generated conversations.

Table~\ref{tab:score_distribution} presents the score distribution for each dimension.


\begin{table}[ht]
\centering
\resizebox{\columnwidth}{!}{%
\begin{tabular}{lcccc}
\toprule
\textbf{Score} & \textbf{Nat.} & \textbf{Coh.} & \textbf{Q. Qual.} & \textbf{Grnd.} \\
\midrule
1 & 0 & 0 & 0 & 0 \\
2 & 0 & 0 & 0 & 3 \\
3 & 39 & 2 & 15 & 19 \\
4 & 387 & 15 & 381 & 146 \\
5 & 281 & 690 & 311 & 539 \\
\bottomrule
\end{tabular}%
}
\caption{Score distribution across four quality dimensions. The concentration of scores at 4 and 5 indicates consistently high quality across the dataset. (Nat.: Naturalness, Coh.: Turn Coherence, Q. Qual.: Question Quality, Grnd.: Groundedness).}
\label{tab:score_distribution}
\end{table}
\subsection{Turn-Level Dependency Analysis}
\label{appendix:dependency_analysis}

Beyond conversation-level quality, we analyzed how each of the 2,971 turns depends on the preceding conversation context. Understanding these dependency patterns is important for conversational retrieval systems, as different dependency types pose different challenges for query understanding and document retrieval.

We classified each turn into one of six dependency types:

\textbf{Coreference}: The question uses pronouns (it, this, that, they, these, those) that point back to entities or concepts mentioned in previous turns. Example: After discussing photosynthesis, asking ``Does it also occur at night?''

\textbf{Ellipsis}: The question is grammatically incomplete and requires prior context to be understood. Example: ``And the second reason?'' or ``What about in winter?''

\textbf{Substitution}: The question uses a general term to refer to something specific mentioned earlier. Example: After explaining a chemical reaction, asking ``How efficient is this process?''

\textbf{Continuation}: The question is grammatically complete and addresses the same topic as previous turns, but without explicit backward references. Example: After discussing photosynthesis mechanisms, asking ``What role does chlorophyll play?''

\textbf{Topic Shift}: The question moves to a new aspect or subtopic of the overall conversation theme. Example: After discussing scientific aspects, asking ``What are the economic implications?''

\textbf{Self-contained}: The question is fully independent and can be understood without any prior context. This typically applies to first-turn questions.

Table~\ref{tab:dependency_distribution} presents the distribution of dependency types across all turns.

\begin{table}[ht]
\centering

\begin{tabular}{lcc}
\toprule
\textbf{Dependency Type} & \textbf{Count} & \textbf{Percentage} \\
\midrule
Continuation & 1,028 & 34.6\% \\
Coreference & 859 & 28.9\% \\
Self-contained & 707 & 23.8\% \\
Topic Shift & 326 & 11.0\% \\
Substitution & 41 & 1.4\% \\
Ellipsis & 10 & 0.3\% \\
\midrule
\textbf{Total} & \textbf{2,971} & \textbf{100\%} \\
\bottomrule
\end{tabular}
\caption{Distribution of turn dependency types. The diversity of dependency patterns reflects natural conversational behavior.}
\label{tab:dependency_distribution}
\end{table}

The dependency type entropy is 0.77 (normalized to 0-1 scale), indicating balanced variety in how turns connect to prior context. The high proportion of continuation (34.6\%) and coreference (28.9\%) turns demonstrates that the generated conversations maintain topical coherence while progressively building on shared context. The self-contained category (23.8\%) aligns closely with the proportion of first turns in the dataset, as expected. Topic shifts (11.0\%) appear at appropriate rates, allowing conversations to explore multiple aspects of complex topics.

\subsection{Turn-Level Question Pattern Analysis}
\label{appendix:question_pattern_analysis}

We also classified each turn by the type of information the question seeks. This analysis reveals the diversity of information-seeking behaviors in the benchmark and ensures that conversations explore topics through varied questioning strategies.

We identified nine question patterns:

\textbf{Why}: Questions asking for reasons or causes. Example: ``Why does this happen?'' or ``What causes this effect?''

\textbf{How}: Questions asking how something works or happens. Example: ``How does the process unfold?'' or ``What is the mechanism?''

\textbf{What}: Questions asking for factual information such as definitions, names, dates, or places. Example: ``What is the definition of X?'' or ``When did this occur?''

\textbf{Compare}: Questions asking about differences or similarities between concepts. Example: ``How does X differ from Y?'' or ``Which approach is more effective?''

\textbf{What-if}: Questions exploring possibilities or hypothetical scenarios. Example: ``What would happen if X changed?'' or ``Could this work under different conditions?''

\textbf{Confirm}: Questions seeking to verify understanding. Example: ``Is it true that X causes Y?'' or ``Does that mean Z?''

\textbf{More-detail}: Questions requesting deeper explanation of something already mentioned. Example: ``Can you explain more about X?'' or ``What specifically happens during Y?''

\textbf{Example}: Questions asking for concrete instances or cases. Example: ``What are some examples of this?'' or ``Can you give a specific case?''

\textbf{Effect}: Questions asking about results, outcomes, or implications. Example: ``What are the effects of X?'' or ``What does this mean for Y?''

Table~\ref{tab:question_pattern_distribution} presents the distribution of question patterns across all turns.

\begin{table}[ht]
\centering

\begin{tabular}{lcc}
\toprule
\textbf{Question Pattern} & \textbf{Count} & \textbf{Percentage} \\
\midrule
How & 571 & 19.2\% \\
Effect & 486 & 16.4\% \\
Confirm & 465 & 15.6\% \\
Why & 432 & 14.5\% \\
Compare & 411 & 13.8\% \\
What-if & 329 & 11.1\% \\
What & 206 & 6.9\% \\
Example & 50 & 1.7\% \\
More-detail & 21 & 0.7\% \\
\midrule
\textbf{Total} & \textbf{2,971} & \textbf{100\%} \\
\bottomrule
\end{tabular}
\caption{Distribution of question patterns. The high entropy indicates diverse information-seeking behavior across conversations.}
\label{tab:question_pattern_distribution}
\end{table}

The question pattern entropy is 0.91 (normalized to 0-1 scale), indicating high diversity in question types. No single pattern dominates the dataset, with the six most frequent patterns (how, effect, confirm, why, compare, what-if) each contributing between 11\% and 19\% of turns. This distribution reflects the reasoning-intensive nature of the benchmark, where users explore topics through causal reasoning (why), procedural understanding (how), consequence analysis (effect), and comparative evaluation (compare), rather than simple factual queries (what, 6.9\%).

\subsection{Per-Domain Results}
\label{appendix:per_domain_results}

Table~\ref{tab:per_domain_results} presents quality scores and turn-level metrics broken down by domain.


\begin{table}[t]
\centering
\resizebox{\columnwidth}{!}{%
\begin{tabular}{@{}lccccccc@{}}
\toprule
\textbf{Dom} & \textbf{\#C} & \textbf{\#T} & \textbf{Nat} & \textbf{Coh} & \textbf{QQ} & \textbf{Grd} & \textbf{DE} \\
\midrule
Biology & 85 & 362 & 4.38 & 4.98 & 4.47 & 4.74 & 0.93 \\
Drones & 37 & 142 & 4.59 & 4.97 & 4.51 & 4.76 & 0.96 \\
Earth Science & 98 & 454 & 4.33 & 5.00 & 4.54 & 4.67 & 0.77 \\
Economics & 74 & 288 & 4.12 & 4.97 & 4.41 & 4.62 & 0.86 \\
Hardware & 46 & 188 & 4.57 & 4.80 & 4.44 & 4.74 & 0.87 \\
Law & 50 & 230 & 4.34 & 5.00 & 4.47 & 4.74 & 0.74 \\
Med. Sciences & 44 & 183 & 4.42 & 4.98 & 4.36 & 4.82 & 0.81 \\
Politics & 43 & 213 & 4.39 & 4.95 & 4.30 & 4.81 & 0.88 \\
Psychology & 84 & 333 & 4.45 & 5.00 & 4.16 & 4.69 & 0.76 \\
Robotics & 68 & 259 & 4.07 & 4.97 & 4.49 & 4.68 & 0.80 \\
Sustain. Living & 78 & 319 & 4.32 & 5.00 & 4.46 & 4.83 & 0.86 \\
\bottomrule
\end{tabular}%
}
\caption{Per-domain quality scores and dependency entropy. \textbf{Dom}: Domain, \textbf{\#C}: No. Conversations, \textbf{\#T}: No. Turns, \textbf{Nat}: Naturalness, \textbf{Coh}: Coherence, \textbf{QQ}: Question Quality, \textbf{Grd}: Grounding, \textbf{DE}: Dependency Entropy.}
\label{tab:per_domain_results}
\end{table}

Quality scores remain consistently high across all eleven domains, with no domain falling below 4.0 on any dimension. Dependency entropy varies from 0.74 (Law) to 0.96 (Drones), reflecting domain-specific conversational patterns. Technical domains such as Drones and Biology exhibit higher dependency diversity, while structured domains such as Law show more uniform turn progression patterns.

\subsection{Evaluation Prompts}
\label{appendix:evaluation_prompts}

This section provides the complete prompts used for automatic evaluation.

\begin{figure*}[!t]
\centering

\begin{tcolorbox}[
  colback=gray!5,
  colframe=gray!75,
  title=Naturalness Evaluation Prompt,
  width=\textwidth
]
\small\ttfamily
Evaluate whether this conversation sounds like NATURAL HUMAN SPEECH.

\textbf{CONVERSATION:}\\
\{conversation\}

\textbf{TASK:} Rate how natural and human-like the language is (not robotic or overly formal).

\textbf{CHECK ONLY these language features:}

\textbf{1. CASUAL WORD CHOICES:}
\begin{itemize}
  \item Contractions used: "don't", "isn't", "what's" (vs formal "do not", "is not")
  \item Casual words: "got", "stuff", "things", "pretty much" (vs formal "obtained", "materials")
  \item Softening words: "kind of", "sort of", "maybe", "I think"
\end{itemize}

\textbf{2. CONVERSATION STARTERS:}
\begin{itemize}
  \item Natural openers: "So", "Well", "Oh", "Yeah", "I mean", "Actually"
  \item Response words: "Right", "Okay", "I see", "Got it"
\end{itemize}

\textbf{3. SENTENCE STYLE:}
\begin{itemize}
  \item Natural incomplete sentences (OK in speech)
  \item Casual question phrasing (not stiff templates)
\end{itemize}

\textbf{DO NOT CHECK (other prompts handle these):}
\begin{itemize}
  \item Whether turns connect logically
  \item Whether questions cover the topic well
  \item Whether answers are factually correct
\end{itemize}

\textbf{SCORING:}\\
5 - Sounds completely natural, like real human speech\\
4 - Mostly natural, few formal spots\\
3 - Mix of natural and formal/stiff language\\
2 - Mostly formal or robotic sounding\\
1 - Entirely artificial, would never occur in real speech

\textbf{Return JSON:}\\
\{\\
~~"score": <1-5>,\\
~~"natural\_phrases": ["quote 2-3 natural-sounding phrases found"],\\
~~"unnatural\_phrases": ["quote any stiff or robotic phrases found"],\\
~~"justification": "1-2 sentences explaining your score"\\
\}
\end{tcolorbox}
\caption{Prompt for evaluating conversation naturalness.}
\label{fig:prompt_naturalness}
\end{figure*}

\begin{figure*}[!t]
\centering

\begin{tcolorbox}[
  colback=gray!5,
  colframe=gray!75,
  title=Turn Coherence Evaluation Prompt,
  width=\textwidth
]
\small\ttfamily
Evaluate whether conversation turns CONNECT LOGICALLY to each other.

\textbf{CONVERSATION:}\\
\{conversation\}

\textbf{TASK:} Rate how well each question follows from the previous answer.

\textbf{CHECK ONLY these connection features:}

\textbf{1. CLEAR REFERENCES:}
\begin{itemize}
  \item When "it", "this", "that", "they" are used, is it clear what they refer to?
  \item When "the problem" or "the process" is mentioned, was it introduced earlier?
\end{itemize}

\textbf{2. LOGICAL CONNECTIONS (how does each question relate to the previous answer?):}
\begin{itemize}
  \item ASKS FOR DETAILS: Question wants more information about something just mentioned
  \item ASKS TO CLARIFY: Question wants to understand something from the answer better
  \item ASKS ABOUT EFFECTS: Question asks what happens as a result of what was described
  \item ASKS FOR CONTRAST: Question asks about an alternative or opposite view
  \item NARROWS FOCUS: Question zooms in on one specific part mentioned
\end{itemize}

\textbf{3. NO GAPS:}
\begin{itemize}
  \item Each question should make sense given what came before
  \item No sudden jumps to unrelated topics without transition
\end{itemize}

\textbf{DO NOT CHECK (other prompts handle these):}
\begin{itemize}
  \item Whether language sounds natural
  \item Whether questions are about different topics
  \item Whether answers are factually correct
\end{itemize}

\textbf{SCORING:}\\
5 - Every question clearly connects to the previous answer\\
4 - Most questions connect well, 1-2 slightly unclear links\\
3 - Some questions connect, but several feel disconnected\\
2 - Many questions don't clearly follow from previous answers\\
1 - Questions seem random, no logical flow

\textbf{Return JSON:}\\
\{\\
~~"score": <1-5>,\\
~~"good\_connections": ["describe 1-2 turns that connect well"],\\
~~"weak\_connections": ["describe any turns that don't connect clearly"],\\
~~"unclear\_references": ["list any 'it/this/that' without clear meaning"],\\
~~"justification": "1-2 sentences explaining your score"\\
\}
\end{tcolorbox}
\caption{Prompt for evaluating turn coherence.}
\label{fig:prompt_coherence}
\end{figure*}

\begin{figure*}[!t]
\centering

\begin{tcolorbox}[
  colback=gray!5,
  colframe=gray!75,
  title=Question Quality Evaluation Prompt,
  width=\textwidth
]
\small\ttfamily
Evaluate whether the questions COVER DIFFERENT USEFUL ASPECTS of the topic.

\textbf{ORIGINAL TOPIC:}\\
\{original\_query\}

\textbf{QUESTIONS:}\\
\{questions\}

\textbf{TASK:} Rate whether the questions explore the topic well without repeating.

\textbf{CHECK ONLY these coverage features:}

\textbf{1. DIFFERENT ASPECTS:}
\begin{itemize}
  \item Does each question ask about something DIFFERENT?
  \item Are any two questions basically asking the same thing in different words?
\end{itemize}

\textbf{2. USEFUL QUESTIONS:}
\begin{itemize}
  \item Are questions specific enough to get clear answers?
  \item Do questions go beyond obvious or trivial information?
  \item Bad examples: "What is X?" (too basic), "Tell me about X" (too vague)
\end{itemize}

\textbf{3. GOOD COVERAGE:}
\begin{itemize}
  \item Together, do the questions explore important parts of the topic?
  \item Are there major aspects of the topic that are missed?
\end{itemize}

\textbf{DO NOT CHECK (other prompts handle these):}
\begin{itemize}
  \item Whether questions sound natural
  \item Whether questions follow logically from answers
  \item Whether the answers are correct
\end{itemize}

\textbf{SCORING:}\\
5 - All questions distinct, specific, and together cover the topic well\\
4 - Most questions distinct and useful, maybe 1 similar pair\\
3 - Some overlap between questions, some too vague, partial coverage\\
2 - Several questions overlap or are too vague/trivial\\
1 - Heavy repetition, mostly vague or useless questions

\textbf{Return JSON:}\\
\{\\
~~"score": <1-5>,\\
~~"aspects\_covered": ["list the different aspects/subtopics the questions address"],\\
~~"repeated\_questions": ["list any question pairs that ask about the same thing"],\\
~~"weak\_questions": ["list any questions that are too vague or trivial"],\\
~~"justification": "1-2 sentences explaining your score"\\
\}
\end{tcolorbox}
\caption{Prompt for evaluating question quality.}
\label{fig:prompt_question_quality}
\end{figure*}

\begin{figure*}[!t]
\centering

\begin{tcolorbox}[
  colback=gray!5,
  colframe=gray!75,
  title=Groundedness Evaluation Prompt,
  width=\textwidth
]
\small\ttfamily
Evaluate whether the answers are SUPPORTED BY THE DOCUMENTS.

\textbf{SOURCE DOCUMENTS:}\\
\{documents\}

\textbf{ANSWERS TO CHECK:}\\
\{answers\}

\textbf{TASK:} Rate whether the answer content can be found in or reasonably inferred from the documents.

\textbf{CHECK ONLY these accuracy features:}

\textbf{1. CLAIMS MATCH DOCUMENTS:}
\begin{itemize}
  \item Can each fact in the answers be found in the source documents?
  \item Are the facts stated correctly (not twisted or misrepresented)?
\end{itemize}

\textbf{2. NOTHING MADE UP:}
\begin{itemize}
  \item Are there any invented names, places, or organizations?
  \item Are there any made-up numbers, dates, or statistics?
  \item Are there any fake technical terms?
\end{itemize}

\textbf{3. NO CONTRADICTIONS:}
\begin{itemize}
  \item Do any claims directly contradict what the documents say?
  \item Are there claims that go way beyond what the documents support?
\end{itemize}

\textbf{ACCEPTABLE:}
\begin{itemize}
  \item Rewording information from documents
  \item Drawing obvious conclusions from stated facts
\end{itemize}

\textbf{NOT ACCEPTABLE:}
\begin{itemize}
  \item Adding facts not found in documents
  \item Stating guesses as if they were facts
\end{itemize}

\textbf{DO NOT CHECK (other prompts handle these):}
\begin{itemize}
  \item Whether answers sound natural
  \item Whether answers connect to each other
  \item Whether the questions were good
\end{itemize}

\textbf{SCORING:}\\
5 - All claims supported by documents, nothing made up\\
4 - Nearly all claims supported (95\%+), only minor inferences\\
3 - Most claims supported (80-95\%), some unsupported but plausible\\
2 - Many claims unsupported (50-80\%), some things made up\\
1 - Mostly unsupported (<50\%), major fabrications

\textbf{Return JSON:}\\
\{\\
~~"score": <1-5>,\\
~~"supported\_claims": ["list 2-3 claims that ARE in the documents"],\\
~~"unsupported\_claims": ["list claims NOT found in documents"],\\
~~"made\_up\_content": ["list any invented facts, names, or numbers"],\\
~~"justification": "1-2 sentences explaining your score"\\
\}
\end{tcolorbox}
\caption{Prompt for evaluating answer groundedness.}
\label{fig:prompt_groundedness}
\end{figure*}

\begin{figure*}[!t]
\centering

\begin{tcolorbox}[
  colback=gray!5,
  colframe=gray!75,
  title=Turn Dependency Classification Prompt,
  width=\textwidth
]
\small\ttfamily
Classify how this question DEPENDS on the previous conversation.

\textbf{PREVIOUS CONVERSATION:}\\
\{prior\_context\}

\textbf{CURRENT QUESTION:}\\
\{current\_question\}

\textbf{TASK:} Pick ONE dependency type that best describes how this question relates to what came before.

\textbf{DEPENDENCY TYPES (pick ONE):}

\textbf{1. "coreference"} - Uses pronouns pointing back to something mentioned before
\begin{itemize}
  \item Look for: it, this, that, they, these, those
  \item Example: "Does IT also affect..." / "What about THAT?"
\end{itemize}

\textbf{2. "ellipsis"} - Incomplete sentence that needs context to understand
\begin{itemize}
  \item Look for: Missing words, fragments like "And...?", "How about...?"
  \item Example: "And the second reason?" / "What about in winter?"
\end{itemize}

\textbf{3. "substitution"} - Uses a general term for something specific mentioned before
\begin{itemize}
  \item Look for: "this process", "that method", "the problem", "such cases"
  \item Example: After "photosynthesis" → "How efficient is this process?"
\end{itemize}

\textbf{4. "continuation"} - Complete question on the same topic, but no explicit links
\begin{itemize}
  \item Look for: Full sentence, related topic, no pronouns pointing back
  \item Example: After photosynthesis → "What role does chlorophyll play?"
\end{itemize}

\textbf{5. "topic\_shift"} - Moves to a new aspect or subtopic
\begin{itemize}
  \item Look for: "What about...", "Regarding...", "Moving to...", new direction
  \item Example: After science → "What are the economic effects?"
\end{itemize}

\textbf{6. "self\_contained"} - Fully independent, makes sense alone (usually Turn 1)
\begin{itemize}
  \item Look for: Could be understood without any prior context
  \item Example: First question of the conversation
\end{itemize}

\textbf{Return JSON:}\\
\{\\
~~"dependency\_type": "one of the 6 types above",\\
~~"evidence": "quote the specific words that show this type",\\
~~"explanation": "one sentence explaining your choice"\\
\}
\end{tcolorbox}
\caption{Prompt for classifying turn dependency types.}
\label{fig:prompt_dependency}
\end{figure*}

\begin{figure*}[!t]
\centering

\begin{tcolorbox}[
  colback=gray!5,
  colframe=gray!75,
  title=Question Pattern Classification Prompt,
  width=\textwidth
]
\small\ttfamily
Classify what TYPE of answer this question is looking for.

\textbf{QUESTION:}\\
\{question\}

\textbf{TASK:} Pick ONE pattern that best describes what kind of information this question wants.

\textbf{QUESTION TYPES (pick ONE):}

\textbf{1. "why"} - Asks for REASONS or CAUSES
\begin{itemize}
  \item Look for: "Why...?", "What causes...?", "What leads to...?"
  \item Wants: Explanations, reasons
\end{itemize}

\textbf{2. "how"} - Asks HOW something WORKS or HAPPENS
\begin{itemize}
  \item Look for: "How does...?", "What is the process...?", "What happens...?"
  \item Wants: Steps, process descriptions
\end{itemize}

\textbf{3. "what"} - Asks for FACTS (definitions, names, dates, places)
\begin{itemize}
  \item Look for: "What is...?", "Who...?", "When...?", "Where...?"
  \item Wants: Specific facts, definitions
\end{itemize}

\textbf{4. "compare"} - Asks about DIFFERENCES or SIMILARITIES
\begin{itemize}
  \item Look for: "How does X differ from Y?", "Is X similar to Y?", "Which is better?"
  \item Wants: Comparisons
\end{itemize}

\textbf{5. "what\_if"} - Asks about POSSIBILITIES or HYPOTHETICALS
\begin{itemize}
  \item Look for: "What would happen if...?", "Could X...?", "What if...?"
  \item Wants: Speculation, possibilities
\end{itemize}

\textbf{6. "confirm"} - Asks to VERIFY understanding
\begin{itemize}
  \item Look for: "Is it true that...?", "Does that mean...?", "So...right?"
  \item Wants: Yes/no confirmation
\end{itemize}

\textbf{7. "more\_detail"} - Asks for MORE INFORMATION on something
\begin{itemize}
  \item Look for: "Can you explain more...?", "What specifically...?", "Tell me more..."
  \item Wants: Deeper explanation
\end{itemize}

\textbf{8. "example"} - Asks for EXAMPLES or INSTANCES
\begin{itemize}
  \item Look for: "What are examples of...?", "Can you give an instance?", "Like what?"
  \item Wants: Concrete cases
\end{itemize}

\textbf{9. "effect"} - Asks about RESULTS or IMPLICATIONS
\begin{itemize}
  \item Look for: "What does this mean for...?", "What are the effects...?", "What happens as a result?"
  \item Wants: Outcomes, impacts
\end{itemize}

\textbf{Return JSON:}\\
\{\\
~~"question\_pattern": "one of the 9 types above",\\
~~"evidence": "quote the specific words that show this type",\\
~~"explanation": "one sentence explaining your choice"\\
\}
\end{tcolorbox}
\caption{Prompt for classifying question patterns.}
\label{fig:prompt_question_pattern}
\end{figure*}

\end{document}